\documentclass[journal,twoside,web]{ieeecolor}
\usepackage{tmi}
\usepackage{cite}
\usepackage{amsmath,amssymb,amsfonts}
\usepackage{algorithmic}
\usepackage{graphicx}
\usepackage{textcomp}

\usepackage{hyperref}       
\usepackage{url}            

\usepackage{subfig}
\usepackage{cleveref}
\usepackage[english]{babel}
\usepackage{multirow, booktabs}
\usepackage{arydshln}

\def\BibTeX{{\rm B\kern-.05em{\sc i\kern-.025em b}\kern-.08em
    T\kern-.1667em\lower.7ex\hbox{E}\kern-.125emX}}
\begin{document}
\title{I$^3$Net: Inter-Intra-slice Interpolation Network for Medical Slice Synthesis}
\author{Haofei Song, Xintian Mao, Jing Yu, Qingli Li, Yan Wang
\thanks{Manuscript submitted on Sep, 2023. This work was supported by the National Natural Science Foundation of China (Grant No. 62101191), Shanghai Natural Science Foundation (Grant No. 21ZR1420800), and the Science and Technology Commission of Shanghai Municipality (Grant No. 22DZ2229004).}
\thanks{H. Song, X. Mao, J. Yu, and Q. Li and Y. Wang are with Shanghai Key Laboratory of Multidimensional Information Processing, East China Normal University, Shanghai, China
(e-mail: hfsong@stu.ecnu.edu.cn, 52265904010@stu.ecnu.edu.cn, 51255904036@stu.ecnu.edu.cn, qlli@cs.ecnu.edu.cn, ywang@cee.ecnu.edu.cn).}
\thanks{Corresponding author: Y. Wang}}

\maketitle

\begin{abstract}
  Medical imaging is limited by acquisition time and scanning equipment. CT and {MR} volumes, reconstructed with thicker slices, are anisotropic with high {in-plane}
  resolution and low {through-plane}
  resolution. We reveal an intriguing phenomenon that due to the mentioned nature of data, performing slice-wise interpolation from the axial view can yield greater benefits than performing super-resolution from other views. Based on this observation, we propose an Inter-Intra-slice Interpolation Network (I$^3$Net), which fully explores information from high {in-plane} resolution and compensates for low {through-plane} resolution. The {through-plane} branch supplements the limited information contained in low {through-plane} resolution from high {in-plane} resolution and enables continual and diverse feature learning. {In-plane} branch transforms features to the frequency domain and enforces an equal learning opportunity for all frequency bands in a global context learning paradigm. We further propose a cross-view block to take advantage of the information from all three views online. Extensive experiments on two public datasets demonstrate the effectiveness of I$^3$Net, and noticeably outperforms state-of-the-art super-resolution, video frame interpolation and slice interpolation methods by a large margin. {We achieve 43.90dB in PSNR, with at least 1.14dB improvement under the upscale factor of $\times$2 on MSD dataset with faster inference.
  Code is available at \url{https://github.com/DeepMed-Lab-ECNU/Medical-Image-Reconstruction}.
  }

\end{abstract}


\begin{figure}[t]
\centering
\subfloat[\label{fig:Fig1.sub.a}]{
        \centering
        \includegraphics[width=.13\textwidth]{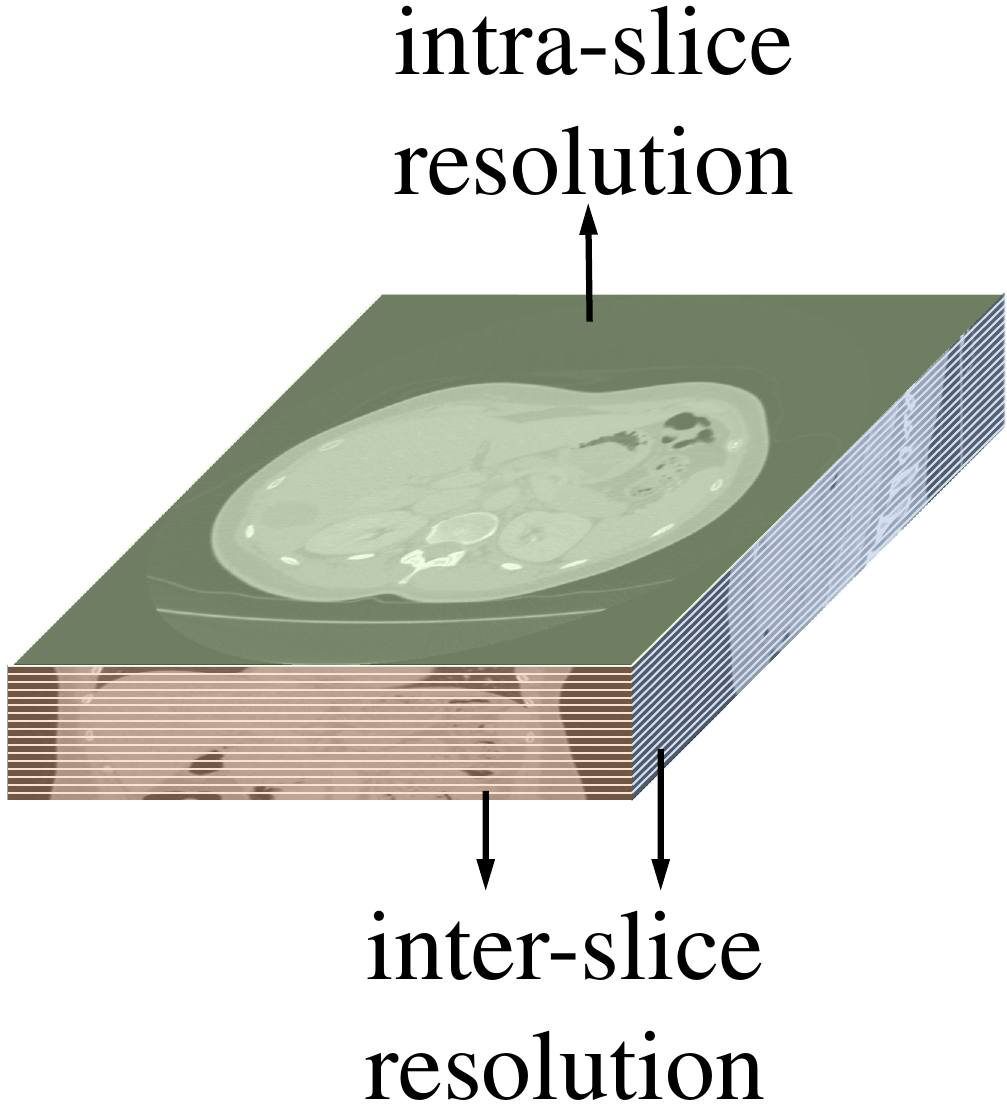}}
\subfloat[\label{fig:Fig1.sub.b}]{
        \includegraphics[width=.34\textwidth]{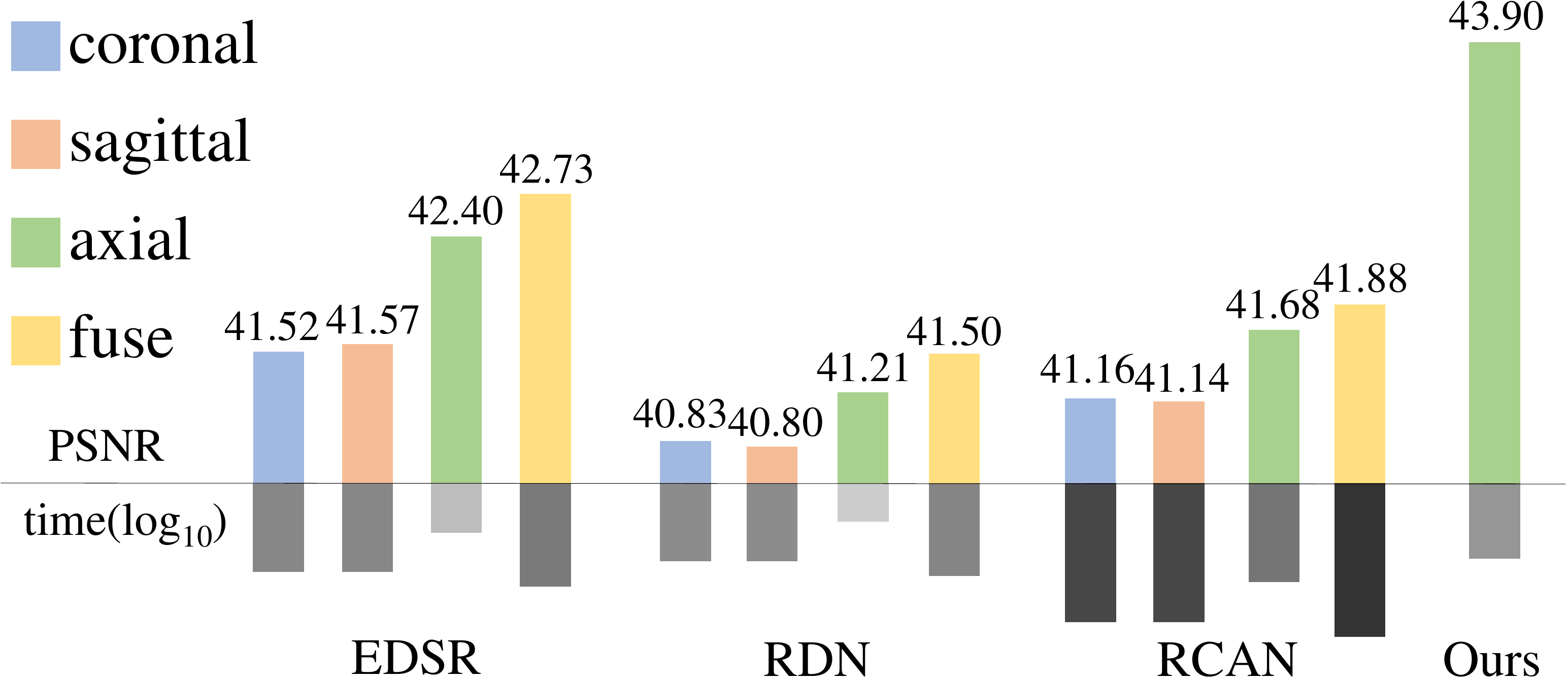}}
\caption{(a) CTs being anisotropic with thick slices. (b) The comparisons of using typical SR backbones from three views and ours. ``coronal'', ``sagittal'' and ``axial'' mean super-resolution from coronal/sagittal view and interpolation from axial view. ``fuse'' means fusing the results from {multiple} views. In contrast, our method directly applies interpolation from the axial view. Evidently, interpolation from the axial view outperforms SR from the other two views. The fusion strategy has a slight improvement, but at a higher cost. We attempt to deal with the medical slice synthesis task via slice-wise interpolation from the axial view. }
\label{fig:Fig1}
\end{figure}

\section{Introduction}
\label{sec:introduction}
\IEEEPARstart{M}{edical} images, such as Computed Tomography (CT) {and Magnetic Resonance (MR)} volume data, provide essential details for disease diagnosis and treatment planning. To suppress noise and reduce the amount of data, some CTs are reconstructed with thicker slices. This results in most CT volumes being anisotropic with high {in-plane} 
resolution and low {through-plane} 
resolution \cite{yu2022rplhr,peng2020saint,fang2022incremental}. {Due to faster acquisitions, MR volumes also have higher in-plane resolution in clinical practice~\cite{peng2020saint,lu2021two}.} Such inconsistent resolutions cause many issues in clinical applications, \emph{e.g.}, losing critical details for lesions, or posing big challenges for designing 3D medical image analysis methods.

Medical slice synthesis is a rising research topic in recent years, which upsamples anisotropic volumes along the low inter-slice resolution to deal with the mentioned problem. But, the nature of the data, \emph{i.e.}, high intra-slice resolution, and low {through-plane} resolution shown in Fig.~\ref{fig:Fig1} (a) have not been fully explored \cite{peng2020saint,yu2022rplhr,fang2022incremental}. In general, two strategies are applied to deal with the slice synthesis problem: (1) slice-wise interpolation from the axial view, which usually considers learning features from the high {in-plane} resolution. {Yu \emph{et al.}~\cite{yu2022rplhr} focuses on recovering in-plane slice based on long-range dependencies.} (2) pixel-wise interpolation (super-resolution) from coronal and sagittal views, which usually considers learning features from the low {through-plane} resolution. {Peng \emph{et al.}~\cite{peng2020saint} directly respectively adopt super-resolution through plane from coronal and sagittal views, and fuse them.} Although a recent work designs both a slice-wise interpolation branch (from axial view) and two pixel-wise interpolation branches (from coronal and sagittal views) to provide complementary knowledge for medical slice synthesis \cite{fang2022incremental}, the global information in high {in-plane} resolution has not been explored, not even in the slice-wise interpolation branch.

We observe an intriguing phenomenon: thanks to the high {in-plane} resolution, the slice-wise interpolation task alone already has the ability to dominate slice synthesis results if {in-plane and through-plane} features are fully excavated. Due to the low {through-plane} resolution, the pixel-wise interpolation task can not make a huge difference. As shown in Fig. \ref{fig:Fig1} (b), we apply three super-resolution methods, \emph{i.e.}, EDSR \cite{lim2017enhanced}, RDN \cite{zhang2018residual} and RCAN \cite{zhang2018image} on the slice-wise interpolation task from the axial view (green bars) and super-resolution tasks from coronal and sagittal views (blue and orange bars). Predictions from three views can be fused to obtain the final results (yellow bars). We also show the inference time in Fig. \ref{fig:Fig1} (b). Slice-wise interpolation from the axial view yields favorable outcomes. The addition of super-resolution tasks from coronal and sagittal views does not result in a significant boost in performance gain in terms of PSNR. But, it does lead to noticeable time costs. This suggests that performing slice-wise interpolation from the axial view can yield greater benefits than performing super-resolution from other views. Thus, instead of equally treating three views and fusing them afterward like \cite{fang2022incremental} or applying super-resolution on coronal and sagittal views like SAINT \cite{peng2020saint}, a more reasonable way is to focus on slice-wise interpolation from the axial view by taking the advantage of high in-plane resolution, and leverage super-resolution from other views as supplementary information. It is worth mentioning that although the use of 3D networks for processing 3D medical images is intuitive \cite{ge2019stereo,sanchez2018brain,zhang2020inter,hatamizadeh2022unetr}, a significant challenge arises due to the large number of parameters.

{To address the above issues, we focus on slice-wise interpolation and propose an Inter-Intra-slice Interpolation Network (I$^3$Net). Specifically, from the perspective of the axial view, inter-slice describes the interaction between adjacent slices along the through-plane direction and intra-slice describes the local variations observed within the slice along the in-plane direction.} While maintaining fast computation speed, it outperforms other methods with three-view fusion, as shown in Fig.~\ref{fig:Fig1} (b). To explore information from high {in-plane} resolution and compensate for low {through-plane} resolution, two branches, an {inter-slice branch} and an {intra-slice branch}, are designed. The inter-slice branch exploits the PixelShuffle operations {~\cite{shi2016real}} before and after feature extraction to supplement the limited information contained in low {through-plane} resolution from high {in-plane} resolution. Since it increases the number of feature channels whose spatial information is similar, it maintains continuity among slices. Features with more channels enable the learning of diverse information. Note that PixelShuffle operations are widely used in {the head/tail of SISR models to reduce the feature size for efficient computation or to reconstruct final outputs}. However, our purpose is to leverage PixelShuffle operations to instantiate continual and diverse feature learning considering the nature of the sparsely-sampled CT. 

Considering that feature learning in the frequency domain can easily learn global context \cite{mao2023intriguing} and different organ contents are emphasized within different frequency ranges in a CT slice, our intra-slice branch divides the features in the frequency domain into non-overlapping windows, and learns features in a cross-frequency and cross-channel manner through MLP-Mixer{~\cite{tolstikhin2021mlp}}. Cross-frequency feature learning enforces an equal learning opportunity for all frequency bands in a global context learning paradigm. Cross-channel feature learning in the frequency domain yields an improved performance with better structural and fine-grained information. 

We further propose a cross-view block to take advantage of the information from all three views. It is worth mentioning that unlike existing methods {\cite{peng2020saint,fang2022incremental}}, which process {different} views separately, our cross-view block extracts complementary information from three views by making features from different views interact with each other in real time without increasing too much computation.

The main contributions can be summarized as follows:
\begin{itemize}
    \item We are the first to look into a reasonable solution for medical slice synthesis by considering the nature of sparsely-sampled CT {and MR} (high {in-plane} resolution and low {through-plane} resolution). 
    \item We propose I$^3$Net with an inter-slice branch and an intra-slice branch. The inter-slice branch compensates for the limited information available in low {through-plane} resolution by utilizing the high {in-plane} resolution. The intra-slice branch ensures equal learning across all frequency bands in a global learning context while improving structural and fine-grained information reconstruction.
    \item We verify the effectiveness of our method on {{Medical Segmentation Decathlon (MSD), KiTS19 and IXI datasets}}. {The PSNR of our method exceeds all state-of-the-arts with at least {1.14dB} improvement on MSD under the upscale factor of $\times$2.}
\end{itemize}

\section{Related Work}

{Due to the requirement for fast acquisition and storage efficiency, CT and MR volume are often characterized by anisotropy with the higher in-plane resolution and the lower through-plane resolution. To improve the through-plane resolution, various approaches can be employed to synthesize slices, including super-resolution from coronal or sagittal views, super-resolution based on implicit neural representation and interpolation from axial view.}

\subsection{Image Super-resolution}

As a basic task in image processing, super-resolution has a large number of research works around this, and various image reconstruction tasks can be borrowed and realized from the method of image super-resolution. Original SISR tasks super-resolve in the spatial dimension. SRCNN\cite{dong2014learning} was the first network proposed for Super-Resolution (SR), followed by a series of classical convolutional network\cite{shi2016real,zhang2018image,zhang2018residual,lim2017enhanced}. 
HAN\cite{niu2020single} introduces holistic attention to explore dependencies across locations, channels, and layers to improve super-resolution performance. With the application of transformer frameworks in image processing, SwinIR\cite{liang2021swinir} was the first designed for super-resolution based on an attention mechanism. {Recently, implicit neural representation (INR) is popular with continuous representation ability to map the coordinates in the latent space. Chen \emph{et al.}\cite{chen2021learning} present a local implicit representation in the self-supervised paradigm, which can reconstruct the extremely high resolution.}
{Sub-pixel convolution layer \cite{shi2016real} is usually used for upsampling at the end of the model. Guo \emph{et al.}~\cite{guo2023asconvsr} adopt PixelUnshuffle and PixelShuffle at the beginning and end of the network as an efficient design. Zhang \emph{et al.}~\cite{zhang2023essaformer} use PixelShuffle and PixelUnshuffle at the beginning and end of each block to enlarge the receptive field.}

Medical slice synthesis can be achieved by image super-resolution from coronal and sagittal views, while only one dimension is enlarged, other dimensions remain unchanged. However, slices from the axial view are not involved in this process. As suggested by our intriguing observation, performing slice interpolation from the axial view is a more reasonable solution for medical slice synthesis.

\begin{figure*}[!t]
    \centering
    \includegraphics[width=\textwidth]{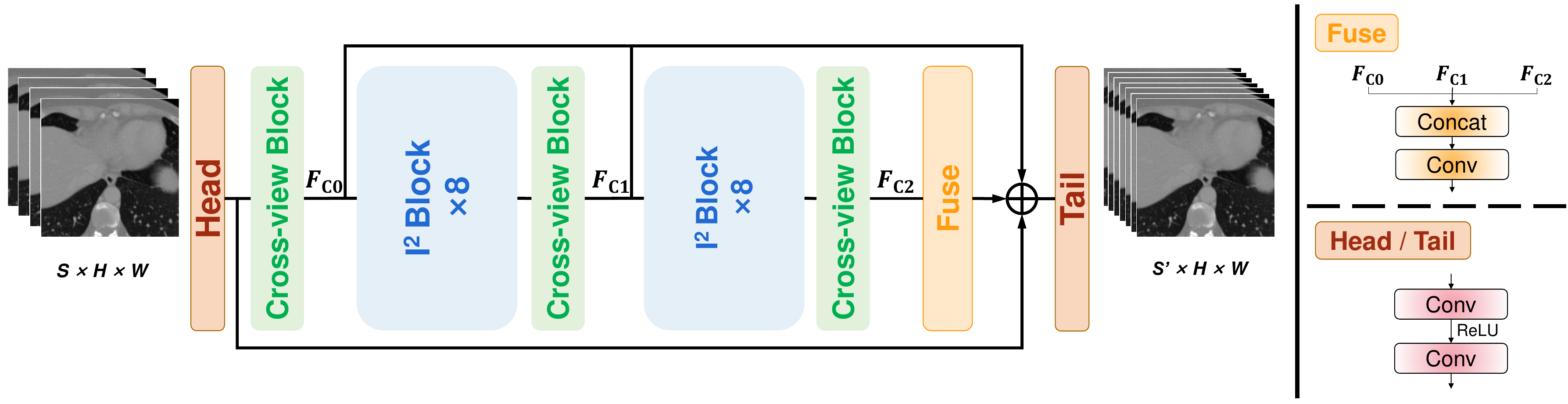}
    \caption{{The architecture of I$^3$Net. Our I$^3$Net consists of several I$^2$ Blocks and three Cross-view Blocks.}}
    \label{fig:architecture}
\end{figure*}

\subsection{Video Frame Interpolation}

Video Frame Interpolation is similar to the slice-wise interpolation. Both tasks involve interpolating frames/slices in the third dimension. But, our task involves interpolating slices in the third spatial dimension, where objects' shape expands or contracts between slices and objects' textures change. There is no explicit correspondence and motion relationship between adjacent slices. For video frame interpolation, the variations observed between consecutive frames primarily arise from the movement of objects within the scene or the motion of the camera capturing the footage. The differences in dynamic changes between frames can be presented by optical flow. Thus, most previous study predicts these differences by estimating optical flow~\cite{wang2019edvr,lin2021fdan}. Other methods interpolate intermediate frames from time-adjacent frames based on adaptive kernel~\cite{niklaus2017video} and flow-based adaptive kernel~\cite{lee2020adacof}.

\subsection{Medical Image Reconstruction}

It is known that medical image has unique properties compared with natural images. Due to balance factors such as sampling time, coat and health risk, there is a trade-off that leads to {a reduction in quality and anisotropy} of sampled medical image. Exclusive research efforts have been tailored towards medical images rather than directly applying methods based on natural images in order to reconstruct these diverse low-quality medical images. {Here are some works on medical image upsampling, including SR and slice synthesis.}
{Sánchez \emph{et al.}~\cite{sanchez2018brain} adopt an adversarial learning approach with 3D convolutions based on the SRGAN~\cite{ledig2017photo} to extract volumetric information. You \emph{et al.}~\cite{you2019ct} leverage a cycle-consistency with the Wasserstein distance for feature mapping in the semi-supervised paradigm. Apart from these methods that explicitly predict each pixel value, there are also some works based on INR. Wu \emph{et al.}~\cite{wu2022arbitrary} achieve arbitrary upsampling by mapping the discrete voxels to a continuous implicit space. McGinnis \emph{et al.}~\cite{mcginnis2023single} integrate two different contrast MR images from different views and establish a continuous spatial function to comprehensively capture the anatomical information. In addition to enhancing in-plane resolution, slice synthesis to improve the through-plane resolution of anisotropic volumes in the axial view is also crucial.} {Bae \emph{et al.}~\cite{bae2018residual} adopt a fully residual convolutional neural network with the input image only from coronal view. Super-resolution from a single view (\emph{i.e.}, coronal and sagittal view) is intuitive, but compared to the axial view, it provides limited information contained in low through-plane resolution. Additionally, it overlooks the anatomical information from other views. Ge \emph{et al.}~\cite{ge2019stereo} propose consecutive 3D multi-scale residual blocks and utilize a stereo-correlation constraint and an image-expression constraint as training guidance.} {Most} current methods conduct super-resolution or interpolation on three views separately and fuse them all. {Peng \emph{et al.}~\cite{peng2020saint} super-resolve from coronal and sagittal views separately, with physical distance between the voxels to provide desirable levels of details, and then fuse them in the second stage.} {Fang \emph{et al.}~\cite{fang2022incremental} design a self-supervised framework based on a cross-view mutual distillation process, which conducts super-resolution from coronal and sagittal views, interpolation from axial view and merging all three views results in the final.} {Yu \emph{et al.}~\cite{yu2022rplhr} apply pure transformer architecture with MAE-like method to increase the number of slices from axial view.} Although some methods~\cite{peng2020saint,fang2022incremental,lu2021two} fuse information from multiple views, they have drawbacks in efficiency. Moreover, when performing multi-view processing, it is crucial to ensure the consistency of fusion across different views.

\subsection{Frequency Domain in Computer Vision}
Frequency transformation transfers images from spatial domain to frequency domain. In this way, each part in frequency domain can map global information into spatial domain. Lama\cite{Suvorov2022resolution} applies Fast Fourier Convolution\cite{chi2020fast} for image inpainting. After that, DeepRFT\cite{mao2023intriguing} proposes Res-FFT-ReLU block for image deblurring because of the ability of global feature extraction. FourierUp~\cite{yu2022deep} rethinks the relationships between spatial and Fourier domains, revealing the transform rules of different resolutions in the Fourier domain.
Xie \emph{et al.}~\cite{Xie2021LearningFD} proposed a frequency-aware dynamic network for SR, which gives different processing resources for different frequency bands.
Zhong \emph{et al.}~\cite{zhong2022detecting} design a frequency enhancement module and fuse feature from frequency and spatial domain with a feature alignment.

\begin{figure}[t]
    \centering
    \includegraphics[width=.5\textwidth]{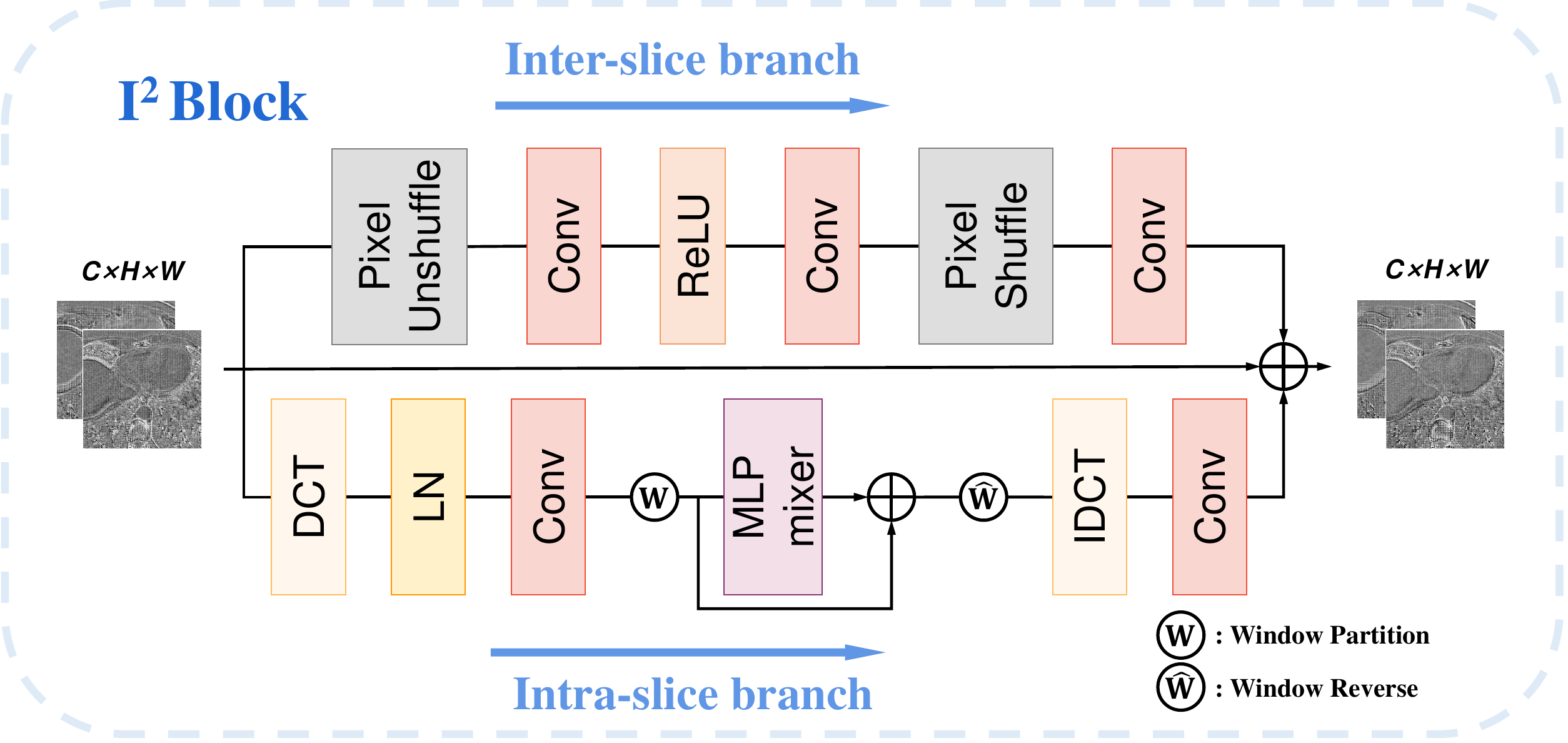}
    \caption{The architecture of I$^2$Block, which consists of an inter-slice branch and an intra-slice branch.}
    \label{fig:architecture_i2block}
\end{figure}

\section{Method}

Mathematically, we define a sparsely-sampled 3D medical image (\emph{e.g.}, CT volume) as $\mathbf{V}\in \mathbb{R}^{S \times H \times W}$, where $S$, $H$ and $W$ indicate the resolution for axial, coronal and sagittal axes. In general, the resolution of the axial axis is lower than the other two axes, \emph{i.e.}, $S<H$, and $S<W$. 
Decomposed from axial, sagittal, and coronal views, the set of obtained 2D images can be denoted as {$\mathbf{V}_{axial}\in \mathbb{R}^{ H \times W}$,$\mathbf{V}_{cor}\in \mathbb{R}^{ S \times W}$,$\mathbf{V}_{sag}\in \mathbb{R}^{ S \times H}$}, respectively.  
The goal of medical slice synthesis is finding a transformation from sparsely-sampled CT volume $\mathbf{V}$ along the axial axis to its densely-sampled CT volume counterpart {$\hat{\mathbf{V}}\in \mathbb{R}^{ \hat{S} \times H \times W}$, where $\hat{S}=(S-1)\cdot R+1$ and} $R$ is the upsampling factor along the axial axis, indicating interpolating $R-1$ slices between every two adjacent slices along the axial axis. 

The overall architecture of the proposed I$^3$Net is shown in Fig.~\ref{fig:architecture}. A CT volume $\mathbf{V}\in\mathbb{R}^{S\times H\times W}$ is fed into the head of the network to produce a feature volume $\mathbf{Z}\in\mathbb{R}^{C\times H\times W}$. Then $\mathbf{Z}$ is fed into several I$^2$ Blocks (Sec.~\ref{sec:I2block}) and Cross-view Blocks (Sec.~\ref{sec:cr}), followed by the tail of the network to generate the interpolated CT volume {$\tilde{\mathbf{V}}\in\mathbb{R}^{\hat{S}\times H\times W}$}. 

\subsection{I$^2$ Block}
\label{sec:I2block}
To extract features from high {in-plane} resolution and compensate for low {through-plane} resolution, we propose an I$^2$ Block, consisting of an intra-slice branch and an inter-slice branch. {The intra-slice branch helps equal learning of all frequency bands to ensure essential high-frequency details when capturing global features}, while the inter-slice branch takes advantage of abundant yet redundant \cite{han2020ghostnet} intra-slice features to compensate for the axial axis.

\subsubsection{Inter-slice Branch}
\begin{figure}[t]
    \centering
    \includegraphics[width=.48\textwidth]{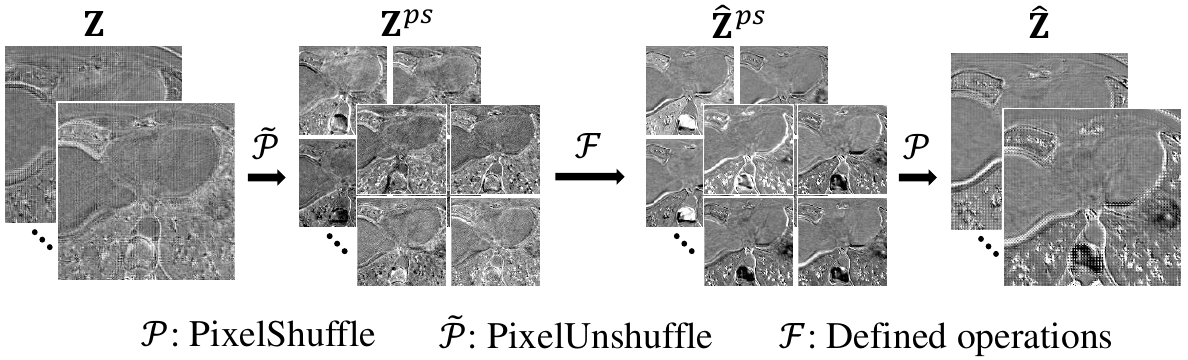}
    \caption{The processing of PixelShuffle and PixelUnshuffle within the inter-slice branch.}
    \label{fig:pixelshuffle}

\end{figure}

As shown in Fig.~\ref{fig:architecture_i2block}, our inter-slice branch consists of {three} types of simple operations: PixelUnshuffle, convolution, ReLU, and PixelShuffle. More specifically, the input feature volume $\mathbf{Z}$ goes through the PixelUnshuffle operation to obtain $\mathbf{Z}^{pu}\in\mathbb{R}^{4C\times H/2\times W/2}$, as illustrated in Fig.~\ref{fig:pixelshuffle}. Thus, the low {through-plane} resolution can be compensated by the high {in-plane} resolution while maintaining continuity. $\mathbf{Z}^{pu}$ enables more information learned for the axial axis via the subsequent convolution-ReLU-convolution operations. PixelShuffle is adopted to transform the feature to its original size. In this branch, we aim at learning enriched and diverse information between different slices.

\subsubsection{Intra-slice Branch}
\label{sec:intra}

{For CT slices from the axial view, low-frequency components always roughly represent the overall morphology of large organs. Small tissues such as blood vessels and airways are blurred and their orientation is difficult to discern in this frequency range. In fact, they are better represented in the high-frequency components.} {This indicates that different organ contents are emphasized within different frequency ranges.}

But, due to the nature of the image, high- and low-frequency information does not have equivalent learning opportunities. The low-frequency part usually dominates the feature extraction process, and in practice, the high-frequency component is typically more challenging to process when learned alongside the lower-frequency component \cite{Xie2021LearningFD}. The fine details in interpolated slices are unsatisfactory, \emph{e.g.}, the blood vessels and airways are out of order, and the contour of organs or spines is missing. Besides, feature learning in the frequency domain can easily acquire global context. Thus, we consider feature learning from the frequency domain.

\begin{figure}[t]
    \centering
    \includegraphics[width=.48\textwidth]{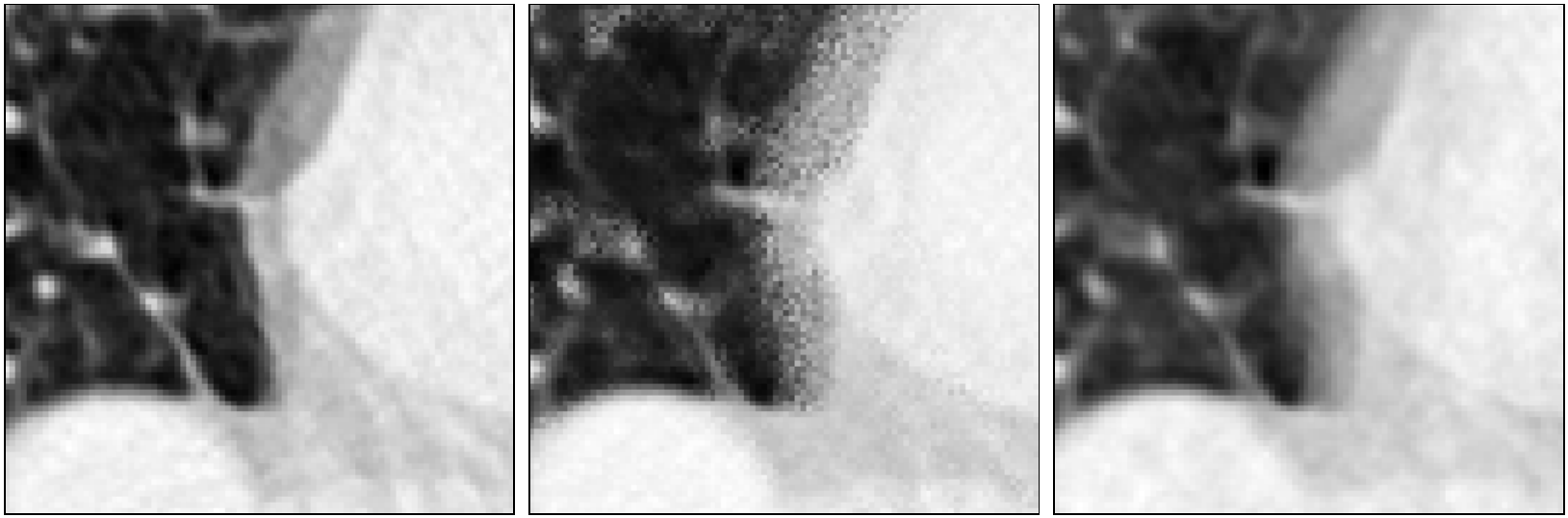}
    \caption{Fusing adjacent slices from spatial domain and spectral domain. The left shows the ground truth of the current slice. The middle is the direct weighted sum of adjacent slices. The right is the weighted sum of the same frequency from adjacent slices.}
    \label{fig:channel_mlp}

\end{figure}

{We employ the Discrete Cosine Transform (DCT) to transform each channel of the input feature $\mathbf{Z}$ from the spatial domain to the frequency domain, and acquire feature volume $\mathbf{Z}^{dct}\in\mathbb{R}^{C\times H\times W}$.}
Since the intensity of the low-frequency region is much greater than that of the high-frequency region, we apply LayerNorm{~\cite{ba2016layer}} on $\mathbf{Z}^{dct}$, which is invariant to per training-case feature shifting and scaling. After that, to enforce an equal learning opportunity for all frequency bands,
we divide the frequency feature along $H\times W$ dimensions into distinct, non-overlapping windows, whose sizes are $C\times p\times p$. Then, the feature in each window can be reshaped to $\mathbf{Z}^{w}_i \in \mathbb{R}^{C \times p^{2}}$ as a sequence of frequency bands with hidden dimension $C$.
In order to further reduce the intensity disparities across different frequencies and enhance the continuity of feature integration, we exploit MLP-Mixer \cite{tolstikhin2021mlp}, \emph{i.e.}, two MLPs including two fully connected layers and an activation function (GELU \cite{hendrycks2016gaussian}). Note that our MLP-Mixer is applied within non-overlapping windows in the frequency domain, which makes it very different from traditional ones. Compared to the global frequencies, the frequencies within the non-overlapping window have comparatively reduced intensity disparities, and still keep the global information. Therefore, one MLP is applied along the frequency dimension, mapping $\mathbb{R}^{p^2}\rightarrow\mathbb{R}^{p^2}$. As shown in Fig.~\ref{fig:channel_mlp}, directly fusing adjacent slices from the spatial domain results in image discretization and indistinctness of tubular structures and edges, while fusing the frequencies of adjacent slices will prevent this situation. So, the other MLP is applied along the channel dimension, mapping $\mathbb{R}^{C}\rightarrow\mathbb{R}^{C}$. In each MLP-Mixer layer, the transformation can be expressed as follows for the input tensor $\mathbf{Z}_i^w$:
\begin{equation}
    {\mathbf{Z}}^{w}_{i,\rm{\uppercase\expandafter{\romannumeral1}}} = \mathbf{Z}^{w}_i +\omega_{2} \sigma (\omega_{1} LN(\mathbf{Z}^{w}_i)),
\end{equation}
\begin{equation}
    {\mathbf{Z}}_{i,\rm{\uppercase\expandafter{\romannumeral2}}}^{w} = {\mathbf{Z}}^{w}_{i,\rm{\uppercase\expandafter{\romannumeral1}}} +\omega_{4} \sigma (\omega_{3} LN({\mathbf{Z}}^{w}_{i,\rm{\uppercase\expandafter{\romannumeral1}}})),
\end{equation}
where {$i = 0,1,...,HW/p^{2}-1$,} $\omega_{1}\sim\omega_4$ are parameters of fully connected layers, $\sigma$ is GELU and $LN$ is LayerNorm. {${\mathbf{Z}}_{i,\rm{\uppercase\expandafter{\romannumeral1}}}^{w}$ and ${\mathbf{Z}}_{i,\rm{\uppercase\expandafter{\romannumeral2}}}^{w}$} represent the outputs after applying the two MLPs, respectively. Then, after applying the window reverse operation to transpose the feature back to size $C\times H\times W$, we utilize $1\times 1$ convolutions to combine the cross-channel context. 

\begin{figure}[t]
    \centering
    \includegraphics[width=.5\textwidth]{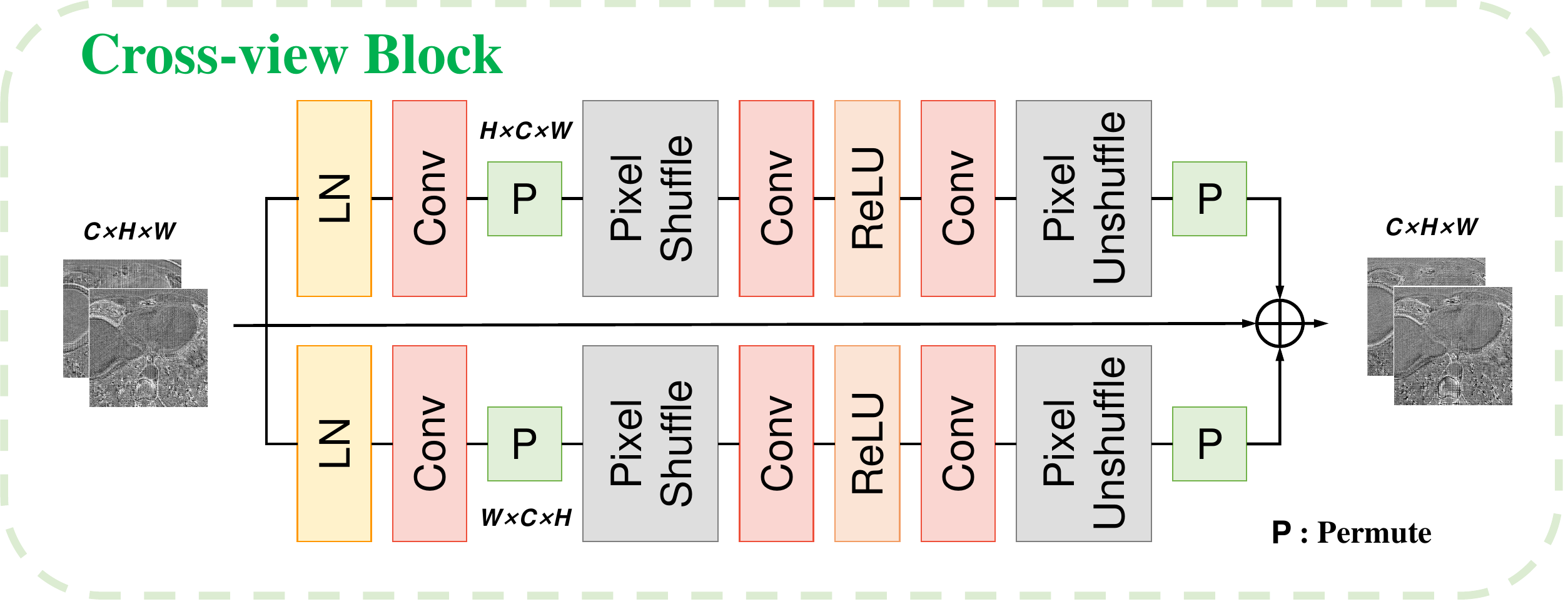}
    \caption{The architecture of the Cross-view Block.}
    \label{fig:architecture_crossviewblock}
\end{figure}

\begin{table*}[t]
    \renewcommand\arraystretch{1.15}
    \centering
    \caption{Comparison with existing SR and interpolation methods under $\times$2, $\times$4 and $\times$6 upsacle factors, on the CT dataset, including colon, liver and hepatic vessels.}
    
    \centering
    \footnotesize
    \setlength{\tabcolsep}{2mm}{
    \resizebox{1\linewidth}{!}{
    \begin{tabular}{|c|l|c|c|c|c| c|c|c|c |c|c|c|c|}
        \hline
        \multicolumn{2}{|c|}{\multirow{2}{*}{Method}} & \multicolumn{4}{c|}{Colon} & \multicolumn{4}{c|}{Liver} & \multicolumn{4}{c|}{Hepatic Vessels} \\ \cline{3-14}
        \multicolumn{2}{|c|}{} & PSNR & SSIM$_a$ & SSIM$_c$ & SSIM$_s$ & PSNR & SSIM$_a$ & SSIM$_c$ & SSIM$_s$ & PSNR & SSIM$_a$ & SSIM$_c$ & SSIM$_s$ \\ \hline
        \multirow{13}{*}{$\times$2} & EDSR2D \cite{lim2017enhanced}   & 42.40 & 0.9841 & 0.9820 & 0.9821 & 41.46 & 0.9806 & 0.9742 & 0.9744 & 43.23 & 0.9858 & 0.9843 & 0.9844 \\
                                    & EDSR3D \cite{lim2017enhanced}   & 42.75 & 0.9845 & 0.9831 & 0.9832 & 41.97 & 0.9817 & 0.9764 & 0.9765 & 43.39 & 0.9860 & 0.9849 & 0.9850 \\
                                    & RDN~\cite{zhang2018residual}    & 41.21 & 0.9811 & 0.9777 & 0.9778 & 40.78 & 0.9783 & 0.9709 & 0.9711 & 41.89 & 0.9830 & 0.9800 & 0.9800 \\
                                    & RCAN~\cite{zhang2018image}     & 41.83 & 0.9827 & 0.9801 & 0.9802 & 41.11 & 0.9769 & 0.9691 & 0.9693 & 43.14 & 0.9856 & 0.9841 & 0.9842 \\
                                    & SwinIR~\cite{liang2021swinir}  & 41.86 & 0.9826 & 0.9802 & 0.9803 & 41.09 & 0.9795 & 0.9727 & 0.9728 & 42.86 & 0.9850 & 0.9834 & 0.9834 \\
                                    & {HAT~\cite{chen2023activating}} & {42.01} & {0.9832} & {0.9809} & {0.9810} & {41.20} & {0.9801} & {0.9732} & {0.9734} & {43.25} & {0.9858} & {0.9845} & {0.9845}\\
                                    & {ArSSR~\cite{wu2022arbitrary}} & {42.73} & {0.9848} & {0.9830} & {0.9831} & {41.75} & {0.9814} & {0.9755} & {0.9756} & {43.59} & {0.9865} & {0.9853} & {0.9853}\\
                                    & AdaCoF~\cite{lee2020adacof}  & 42.49 & 0.9845 & 0.9827 & 0.9828 & 41.65 & 0.9819 & 0.9759 & 0.9761 & 42.77 & 0.9850 & 0.9832 & 0.9833 \\ 
                                    & {RSTT~\cite{geng2022rstt}} & {42.09} & {0.9831} & {0.9813} & {0.9814} & {40.98} & {0.9788} & {0.9720} & {0.9722} & {43.28} & {0.9858} & {0.9846} & {0.9847}\\
                                    & IFRNet~\cite{kong2022ifrnet} & 42.68 & 0.9854 & 0.9834 & 0.9835 & 41.85 & 0.9821 & 0.9769 & 0.9771 & 42.88 & 0.9856 & 0.9838 & 0.9838 \\
                                    & TVSRN~\cite{yu2022rplhr}     & 41.37 & 0.9809 & 0.9782 & 0.9783 & 40.37 & 0.9755 & 0.9681 & 0.9683 & 42.28 & 0.9834 & 0.9812 & 0.9814 \\
                                    & SAINT~\cite{peng2020saint}   & 42.76 & 0.9845 & 0.9836 & 0.9837 & 42.24 & 0.9823  & 0.9784  & 0.9785  & 43.35 & 0.9858 & 0.9850 & 0.9851 \\ \cline{2-14}
                                    & \textbf{I$^3$3Net (ours)}    & \textbf{43.90} & \textbf{0.9871} & \textbf{0.9863} & \textbf{0.9864} & \textbf{42.68} & \textbf{0.9838} & \textbf{0.9796} & \textbf{0.9798} & \textbf{44.60} & \textbf{0.9883} & \textbf{0.9877} & \textbf{0.9878} \\ \hline
        \multirow{12}{*}{$\times$4} & EDSR2D \cite{lim2017enhanced}   & 35.63 & 0.9544 & 0.9939 & 0.9395 & 35.18 & 0.9499 & 0.9232 & 0.9236 & 36.50 & 0.9592 & 0.9461 & 0.9465 \\
                                    & EDSR3D \cite{lim2017enhanced}   & 35.85 & 0.9551 & 0.9422 & 0.9426 & 35.43 & 0.9514 & 0.9265 & 0.9270 & 35.46 & 0.9529 & 0.9362 & 0.9368 \\
                                    & RDN~\cite{zhang2018residual}    & 34.56 & 0.9469 & 0.9269 & 0.9274 & 34.55 & 0.9444 & 0.9152 & 0.9157 & 35.48 & 0.9437 & 0.9355 & 0.9361 \\
                                    & RCAN~\cite{zhang2018image}      & 35.08 & 0.9510 & 0.9331 & 0.9330 & 34.65 & 0.9458 & 0.9165 & 0.9170 & 35.55 & 0.9540 & 0.9361 & 0.9367 \\
                                    & SwinIR~\cite{liang2021swinir}   & 35.32 & 0.9522 & 0.9361 & 0.9365 & 34.98 & 0.9483 & 0.9212 & 0.9216 & 36.48 & 0.9588 & 0.9462 & 0.9467 \\
                                    & {HAT~\cite{chen2023activating}} & {35.61} & {0.9540} & {0.9392} & {0.9395} & {35.17} & {0.9507} & {0.9240} & {0.9243} & {37.00} & {0.9618} & {0.9511} & {0.9515}\\
                                    & {ArSSR~\cite{wu2022arbitrary}} & {35.77} & {0.9551} & {0.9405} & {0.9409} & {35.28} & {0.9504} & {0.9242} & {0.9247} & {36.83} & {0.9612} & {0.9496} & {0.9500}\\
                                    & {RSTT~\cite{geng2022rstt}} & {35.35} & {0.9517} & {0.9364} & {0.9369} & {34.81} & {0.9461} & {0.9187} & {0.9192} & {36.71} & {0.9598} & {0.9485} & {0.9489}\\
                                    & IFRNet~\cite{kong2022ifrnet}    & 36.46 & 0.9612 & 0.9495 & 0.9497 & 35.92 & 0.9552 & 0.9342 & 0.9346 & 36.65 & 0.9616 & 0.9504 & 0.9507 \\
                                    & TVSRN~\cite{yu2022rplhr}        & 35.75 & 0.9541 & 0.9404 & 0.9407 & 35.01 & 0.9483 & 0.9210 & 0.9214 & 36.85 & 0.9606 & 0.9498 & 0.9502 \\ 
                                    & SAINT~\cite{peng2020saint}      & 35.87 & 0.9552 & 0.9440 & 0.9442 & 35.95 & 0.9550 &  0.9355 & 0.9358 & 35.70 & 0.9560 & 0.9423 & 0.9423 \\ \cline{2-14}
                                    & \textbf{I$^{3}$Net (ours)}   & \textbf{37.21} & \textbf{0.9632} & \textbf{0.9545} & \textbf{0.9547} & \textbf{36.31} & \textbf{0.9575} & \textbf{0.9369} & \textbf{0.9373} & \textbf{37.90} & \textbf{0.9670} & \textbf{0.9588} & \textbf{0.9592} \\  \hline

        \multirow{12}{*}{$\times$6}  & EDSR2D~\cite{lim2017enhanced}  & 33.09 & 0.9352 & 0.9091 & 0.9100 & 32.70 & 0.9310 & 0.8889 & 0.8896 & 34.15 & 0.9443 & 0.9198 & 0.9208 \\
                                    & EDSR3D~\cite{lim2017enhanced}  & 32.71 & 0.9320 & 0.9062 & 0.9071 & 32.85 & 0.9321 & 0.8916 & 0.8924 & 32.94 & 0.9343 & 0.9043 & 0.9054 \\
                                    & RDN~\cite{zhang2018residual}   & 32.24 & 0.9270 & 0.8964 & 0.8974 & 32.13 & 0.9245 & 0.8792 & 0.8800 & 33.28 & 0.9385 & 0.9085 & 0.9097 \\
                                    & RCAN~\cite{zhang2018image}  & 33.12 & 0.9358 & 0.9102 & 0.9111 & 32.03 & 0.9237 & 0.8774 & 0.8783 & 33.43 & 0.9398 & 0.9102 & 0.9114 \\
                                    & SwinIR~\cite{liang2021swinir}  & 32.79 & 0.9317 & 0.9050 & 0.9059 & 32.60 & 0.9298 & 0.8873 & 0.8880 & 34.18 & 0.9443 & 0.9208 & 0.9218 \\
                                    & {HAT~\cite{chen2023activating}} & {33.03} & {0.9341} & {0.9088} & {0.9096} & {32.77} & {0.9322} & {0.8904} & {0.8911} & {34.70} & {0.9479} & {0.9279} & {0.9288}\\
                                    & {ArSSR~\cite{wu2022arbitrary}} & {33.13} & {0.9355} & {0.9101} & {0.9110} & {32.74} & {0.9316} & {0.8898} & {0.8905} & {34.32} & {0.9455} & {0.9234} & {0.9243}\\
                                    & {RSTT~\cite{geng2022rstt}} & {32.94} & {0.9331} & {0.9077} & {0.9086} & {32.71} & {0.9311} & {0.8899} & {0.8906} & {34.62} & {0.9472} & {0.9273} & {0.9282}\\
                                    & IFRNet~\cite{kong2022ifrnet}   & 33.88 & 0.9443 & 0.9231 & 0.9237 & 33.23 & 0.9374 & 0.9010 & 0.9016 & 34.20 & 0.9245 & 0.9245 & 0.9245 \\
                                    & TVSRN~\cite{yu2022rplhr}       & 33.28 & 0.9355 & 0.9122 & 0.9130 & 32.84 & 0.9321 & 0.8910 & 0.8921 & 34.62 & 0.9471 & 0.9274 & 0.9283 \\
                                    & SAINT~\cite{peng2020saint}     & 32.76 & 0.9327 & 0.9099 & 0.9107 & 33.26 & 0.9366 & 0.9029 & 0.9034 & 33.03 & 0.9376 & 0.9116 & 0.9126 \\ \cline{2-14}
                                    & \textbf{I$^{3}$Net (ours)}     & \textbf{34.42} & \textbf{0.9459} & \textbf{0.9291} & \textbf{0.9296} & \textbf{33.84} & \textbf{0.9406} & \textbf{0.9067} & \textbf{0.9072} & \textbf{35.28} & \textbf{0.9530} & \textbf{0.9365} & \textbf{0.9372} \\  \hline
                                    
        \end{tabular} %
            }
    }
    \label{tab:comparison_MSD}
\end{table*}

\subsection{Cross-view Block}
\label{sec:cr}
As shown in Fig.~\ref{fig:Fig1}, the axial view contains much more information than other views. Our interpolation network processes slices from the axial view. 
But, sometimes even if the cross-section of the spine is realistic and complete from each slice in the axial view, the morphological results of the spine from the coronal or sagittal view may be off the correct orientation. Therefore, it is necessary to take advantage of the information from all three views and refine the features learned by the axial view. Thus, the morphological position of the organ will be corrected in the interpolated slice. Existing methods \cite{peng2020saint,fang2022incremental} process three views separately, and then followed by feature fusion in the final operation, where these features do not interact with each other in real time. 

\begin{table*}[t]
    \renewcommand\arraystretch{1.15}
    \centering
    \caption{{Comparison with existing SR and interpolation methods under $\times$2, $\times$4 and $\times$6 upsacle factors, on IXI dataset, including three modalities (T1,T2 and PD).}}
    
    \centering
    \footnotesize
    \setlength{\tabcolsep}{2mm}{
    \resizebox{1\linewidth}{!}{
    \begin{tabular}{|c|l|c|c|c|c| c|c|c|c |c|c|c|c|}
        \hline
        \multicolumn{2}{|c|}{\multirow{2}{*}{Method}} & \multicolumn{4}{c|}{T1} & \multicolumn{4}{c|}{T2} & \multicolumn{4}{c|}{PD} \\ \cline{3-14}
        \multicolumn{2}{|c|}{} & PSNR & SSIM$_a$ & SSIM$_c$ & SSIM$_s$ & PSNR & SSIM$_a$ & SSIM$_c$ & SSIM$_s$ & PSNR & SSIM$_a$ & SSIM$_c$ & SSIM$_s$ \\ \hline
        \multirow{13}{*}{$\times$2} & EDSR2D \cite{lim2017enhanced}   & 47.59 & 0.9912 & 0.9914 & 0.9913 & 49.75 & 0.9938 & 0.9935 & 0.9937 & 47.41 & 0.9925 & 0.9919 & 0.9922 \\
                                    & EDSR3D \cite{lim2017enhanced}   & 47.95 & 0.9918 & 0.9920 & 0.9919 & 50.08 & 0.9940 &	0.9938 & 0.9940 & 47.52 & 0.9915 & 0.9915 &	0.9918 \\
                                    & RDN~\cite{zhang2018residual}    & 47.10 & 0.9902 & 0.9904 & 0.9903 & 49.53 & 0.9935 &	0.9932 & 0.9935 & 47.16 & 0.9921 & 0.9915 &	0.9918 \\
                                    & RCAN~\cite{zhang2018image}     & 47.35 & 0.9907 &	0.9909 & 0.9908 & 49.63 & 0.9937 & 0.9934 &	0.9936 & 47.23 & 0.9922 & 0.9916 & 0.9920 \\
                                    & SwinIR~\cite{liang2021swinir}  & 47.35 & 0.9907 &	0.9909 & 0.9908 & 49.65 & 0.9937 & 0.9934 &	0.9936 & 47.25 & 0.9923 & 0.9917 & 0.9921 \\
                                    & HAT~\cite{chen2023activating} & 47.44 & 0.9909 & 0.9911 &	0.9910 & 49.78 & 0.9938 & 0.9936 & 0.9938 & 47.43 &	0.9924 & 0.9919 & 0.9922 \\
                                    & ArSSR~\cite{wu2022arbitrary} & 47.80 & 0.9916 & 0.9918 & 0.9917 & 49.96 & 0.9940 & 0.9937 & 0.9939 & 47.46 & 0.9926 & 0.9921 & 0.9924  \\
                                    & AdaCoF~\cite{lee2020adacof}  & 47.34 & 0.9907 & 0.9909 & 0.9908 & 49.25 &	0.9934 & 0.9930 & 0.9932 & 46.99 & 0.9921 &	0.9915 & 0.9918 \\ 
                                    & RSTT~\cite{geng2022rstt}     & 47.12 & 0.9903 & 0.9905 & 0.9903 & 49.43 & 0.9935 & 0.9932 & 0.9934 & 46.92 & 0.9919 &	0.9912 & 0.9916 \\
                                    & IFRNet~\cite{kong2022ifrnet} & 47.52 & 0.9910 & 0.9913 & 0.9912 & 49.40 &	0.9935 & 0.9931 & 0.9934 & 46.97 & 0.9921 &	0.9914 & 0.9917 \\
                                    & TVSRN~\cite{yu2022rplhr}     & 46.60 & 0.9887 & 0.9889 & 0.9888 & 49.29 &	0.9928 & 0.9925 & 0.9928 & 46.93 & 0.9915 &	0.9908 & 0.9912 \\
                                    & SAINT~\cite{peng2020saint}   & 47.17 & 0.9904 & 0.9906 & 0.9904 & 50.10 &	0.9941 & 0.9939 & 0.9941 & 47.59 & 0.9927 & 0.9923 & 0.9926 \\ \cline{2-14}
                                    & \textbf{I$^3$3Net (ours)}    & \textbf{48.26} & \textbf{0.9923} & \textbf{0.9926} & \textbf{0.9925} & \textbf{50.25} & \textbf{0.9943} & \textbf{0.9940} & \textbf{0.9942} & \textbf{47.79} & \textbf{0.9929} & \textbf{0.9925} & \textbf{0.9928} \\ \hline
        \multirow{12}{*}{$\times$4} & EDSR2D \cite{lim2017enhanced}   & 41.83 &	0.9708 & 0.9702 & 0.9701 & 43.39 & 0.9787 & 0.9773 & 0.9778 & 41.18 & 0.9751 & 0.9732 &	0.9738 \\
                                    & EDSR3D \cite{lim2017enhanced}   & 42.35 &	0.9732 & 0.9733 & 0.9733 & 43.90 & 0.9802 & 0.9792 & 0.9797 & 41.52 & 0.9765 & 0.9749 &	0.9755 \\
                                    & RDN~\cite{zhang2018residual}    & 41.09 &	0.9659 & 0.9646 & 0.9645 & 42.95 & 0.9771 & 0.9754 & 0.9759 & 40.77 & 0.9734 & 0.9710 &	0.9717 \\
                                    & RCAN~\cite{zhang2018image}      & 41.58 &	0.9693 & 0.9684 & 0.9682 & 43.22 & 0.9782 & 0.9767 & 0.9771 & 41.05 & 0.9747 & 0.9725 &	0.9732 \\
                                    & SwinIR~\cite{liang2021swinir}   & 41.59 &	0.9693 & 0.9686 & 0.9685 & 43.11 & 0.9778 & 0.9763 & 0.9767 & 40.95 & 0.9741 & 0.9721 &	0.9727 \\
                                    & HAT~\cite{chen2023activating}   & 41.74 &	0.9703 & 0.9697 & 0.9696 & 43.81 & 0.9802 & 0.9791 & 0.9796 & 41.05 & 0.9746 & 0.9726 &	0.9732 \\
                                    & ArSSR~\cite{wu2022arbitrary}    & 41.86 & 0.9710 & 0.9704 & 0.9704 & 43.73 & 0.9798 & 0.9787 & 0.9792 & 41.28 & 0.9757 & 0.9739 & 0.9745 \\
                                    & RSTT~\cite{geng2022rstt}        & 41.28 &	0.9673 & 0.9663 & 0.9662 & 43.09 & 0.9777 & 0.9761 & 0.9766 & 40.76 & 0.9732 & 0.9709 &	0.9716 \\
                                    & IFRNet~\cite{kong2022ifrnet}    & 42.65 &	0.9749 & 0.9749 & 0.9749 & 43.70 & 0.9796 & 0.9784 & 0.9789 & 41.30 & 0.9757 & 0.9739 &	0.9745 \\
                                    & TVSRN~\cite{yu2022rplhr}        & 41.53 &	0.9689 & 0.9681 & 0.9680 & 43.19 & 0.9776 & 0.9762 & 0.9767 & 40.92 & 0.9739 & 0.9719 &	0.9726 \\ 
                                    & SAINT~\cite{peng2020saint}      & 41.45 &	0.9686 & 0.9676 & 0.9675 & 43.14 & 0.9780 & 0.9765 & 0.9770 & 40.87 & 0.9742 & 0.9720 &	0.9727 \\ \cline{2-14}
                                    & \textbf{I$^{3}$Net (ours)}   & \textbf{43.03} & \textbf{0.9765} & \textbf{0.9768} & \textbf{0.9768} & \textbf{44.39} & \textbf{0.9818} & \textbf{0.9810} & \textbf{0.9814} & \textbf{42.02} & \textbf{0.9783} & \textbf{0.9770} & \textbf{0.9776} \\  \hline

        \multirow{12}{*}{$\times$6} & EDSR2D~\cite{lim2017enhanced}  & 39.31 & 0.9519 &	0.9495 & 0.9495 & 41.40 & 0.9699 & 0.9679 &	0.9685 & 39.48 & 0.9664 & 0.9639 & 0.9648 \\
                                    & EDSR3D~\cite{lim2017enhanced}  & 39.96 & 0.9564 &	0.9556 & 0.9556 & 42.18 & 0.9735 & 0.9723 &	0.9729 & 40.15 & 0.9695 & 0.9680 & 0.9689 \\
                                    & RDN~\cite{zhang2018residual}   & 38.54 & 0.9439 &	0.9403 & 0.9402 & 40.63 & 0.9655 & 0.9628 &	0.9633 & 38.82 & 0.9621 & 0.9587 & 0.9595 \\
                                    & RCAN~\cite{zhang2018image}     & 39.19 & 0.9505 &	0.9479 & 0.9479 & 41.30 & 0.9693 & 0.9673 &	0.9678 & 39.41 & 0.9657 & 0.9631 & 0.9640 \\
                                    & SwinIR~\cite{liang2021swinir}  & 39.22 & 0.9507 &	0.9484 & 0.9484 & 41.14 & 0.9684 & 0.9664 &	0.9669 & 39.20 & 0.9645 & 0.9618 & 0.9626 \\
                                    & HAT~\cite{chen2023activating}  & 39.46 & 0.9528 &	0.9510 & 0.9510 & 41.35 & 0.9696 & 0.9677 &	0.9683 & 39.50 & 0.9663 & 0.9641 & 0.9650 \\
                                    & ArSSR~\cite{wu2022arbitrary}   & 39.37 & 0.9523 & 0.9502 & 0.9502 & 41.81 & 0.9721 & 0.9705 & 0.9710 & 39.66 & 0.9673 & 0.9651 & 0.9660 \\
                                    & RSTT~\cite{geng2022rstt}       & 39.33 & 0.9519 &	0.9498 & 0.9498 & 41.51 & 0.9705 & 0.9688 &	0.9694 & 39.18 & 0.9645 & 0.9617 & 0.9627 \\
                                    & IFRNet~\cite{kong2022ifrnet}   & 40.73 & 0.9625 &	0.9621 & 0.9622 & 42.44 & 0.9748 & 0.9735 &	0.9741 & 40.32 & 0.9708 & 0.9691 & 0.9700 \\
                                    & TVSRN~\cite{yu2022rplhr}       & 39.52 & 0.9531 &	0.9514 & 0.9514 & 41.70 & 0.9713 & 0.9698 &	0.9704 & 39.67 & 0.9664 & 0.9642 & 0.9651 \\
                                    & SAINT~\cite{peng2020saint}     & 38.86 & 0.9474 &	0.9444 & 0.9444 & 40.83 & 0.9674 & 0.9641 &	0.9647 & 38.88 & 0.9629 & 0.9589 & 0.9597 \\ \cline{2-14}
                                    & \textbf{I$^{3}$Net (ours)}     & \textbf{40.84} & \textbf{0.9628} & \textbf{0.9627} & \textbf{0.9628} & \textbf{43.35} & \textbf{0.9786} & \textbf{0.9780} & \textbf{0.9785} & \textbf{41.24} & \textbf{0.9748} & \textbf{0.9739} & \textbf{0.9747} \\ \hline
                                    
        \end{tabular} %
            }
    }
    \label{tab:comparison_IXI}
\end{table*}

For this reason, we design a cross-view block shown in Fig. \ref{fig:architecture_crossviewblock} to unite the information of three views simultaneously in one step within a single network. 
Given features $\mathbf{Z}^{axial} \in \mathbb{R}^{C \times H \times W} $ from the axial view, $\mathbf{Z}^{norm}$ is obtained from 
$\mathbf{Z}^{axial}$ through LayerNorm. Features from the sagittal view $\mathbf{Z}^{sag}$ are extracted along the channel and height dimensions, while features from coronal view $\mathbf{Z}^{cor}$ are extracted along the channel and width dimensions. The process of cross-view block can be represented as follows and $\mathbf{Z}^{enhance}$ is obtained as the output:
\begin{equation}
    \mathbf{Z}^{sag}_{*,*,i} = \text{Down}(\text{ConvUnit}(\text{Up} (\mathbf{W}^{sag} \mathbf{Z}^{norm}_{*,*,i}))), 
\end{equation}
\begin{equation}
    \mathbf{Z}^{cor}_{*,j,*} = \text{Down}(\text{ConvUnit}(\text{Up} (\textbf{W}^{cor} \mathbf{Z}^{norm}_{*,j,*}))), 
\end{equation}
\begin{equation}
    \mathbf{Z}^{enhance} = \mathbf{Z}^{sag} + \mathbf{Z}^{cor} + \mathbf{Z}^{axial},
\end{equation}
where $i=1,...,W$ and $j=1,...,H$. $\mathbf{W}^{sag}$ and $\mathbf{W}^{cor}$ are two $1 \times 1$ convolution operations. $\text{ConvUnit}$ indicates {Conv$3\times 3$-ReLU-Conv$3\times 3$}. $\text{Up}$ and $\text{Down}$ are PixelShuffle and PixelUnshuffle, respectively.

\subsection{Optimization}
Following prior works~\cite{lim2017enhanced,peng2020saint,yu2022rplhr}, we adopt a standard $L_{1}$ Loss between prediction $\tilde{\mathbf{V}}$ and ground truth $\hat{\mathbf{V}}$ as follows:
\begin{equation}
    \mathcal{L} = \left \| \hat{\mathbf{V}} -\tilde{\mathbf{V}} \right \| _{1}
\end{equation}

\section{Experiments}

\subsection{Experiments setting}

\subsubsection{Datasets}
{Both CT and MR are considered in our experiments}. The 3D CT volumes are collected from Medical Segmentation Decathlon (MSD)  \cite{antonelli2022medical}, which is publicly available. The datasets we use contain 201, 433, 190 volumes of liver, hepatic vessels and colon, respectively. We follow the setting in \cite{peng2020saint} to split the dataset for training and testing. 
All the volumes have the size of  $512 \times 512$ within the axial-view slice, and the number of slices per volume ranges from 24 to 1046. The {in-plane} resolution ranges from 0.58mm to 1.0mm, while the {through-plane} resolution ranges from 0.8mm to 8.0mm. It is evident that the 3D CT volume exhibits anisotropy, with generally poor {through-plane} resolution. Furthermore, following \cite{peng2020saint}, in order to verify the cross-dataset generalization ability of our model, we select 32 kidney volumes from the 2019 Kidney Tumor Segmentation Challenge (KiTS19)~\cite{heller2019kits19} as the test set. These volumes, in comparison to those from MSD, have higher {in-plane} resolution.
{The 3D MR volumes are collected from public dataset IXI\footnote{\url{https://brain-development.org/ixi-dataset/}}, containing three modalities (PD, T1 and T2). We choose 185 volumes from Hammersmith Hospital scanned by a Philips 3T system, which are split into 150 volumes for training and 35 volumes for testing. The in-plane dimensions of each volume are $256 \times 256$, with the majority of volumes containing 130 slices. Almost all volumes have the in-plane resolution of 0.9375mm and the through-plane resolution of 1.25mm. }

In our experiment, following \cite{peng2020saint}, we set the downsampling factor along the axial axis as $R=2,4,6$. The original volume is considered as the high-resolution volume, while the low-resolution volume is obtained through direct downsampling of the original volume.

\subsubsection{Implementation Details}
The framework is implemented in PyTorch. The model is optimized by Adam \cite{kingma2014adam} and trained for 1500 epochs with a batch size of 6, using NVIDIA GeForce RTX 3090. The learning rate is initially set to 3e-4 and adjusted by a cosine learning rate decay scheduler. We center crop $256 \times 256$ region in each slice and use it for training and testing. 
To effectively utilize all training data, we adopt a strategy of training on 4 consecutive slices randomly sampled from each training volume, which means $256 \times 256 \times 4$ cubes serve as the low-resolution patches. While the corresponding ground truth are $ 256 \times 256 \times S$ , where $ S = (4-1) \times R + 1$ . In order to address the issue of excessive memory consumption, the low-resolution patches are set to the size of $64\times 64\times 16$ for 3D CNN network. 

During training, we randomly sample 4 slices from each volume within one epoch, while comparison models such as \cite{lee2020adacof,kong2022ifrnet,peng2020saint} require {traversing all slices within each volume,} which takes {more training resources} than ours. To account for this, we train their models on both our setting and their default setting, and report their results with the \textbf{BEST} performance.

\subsubsection{Evaluate Metrics}
Two metrics are applied to measure the interpolated results quality, including Peak Signal-to-Noise Ratio (PSNR) and Structural Similarity Index (SSIM). SSIM is calculated on slices from each view independently, where SSIM$_\textrm{a}$, SSIM$_\textrm{c}$, SSIM$_\textrm{s}$ are obtained from axial, coronal and sagittal views, respectively.

\subsection{Comparison between I$^3$Net and Other Methods}
\subsubsection{Performance Comparison}

\begin{figure*}
	\centering
    \includegraphics[width=\linewidth]{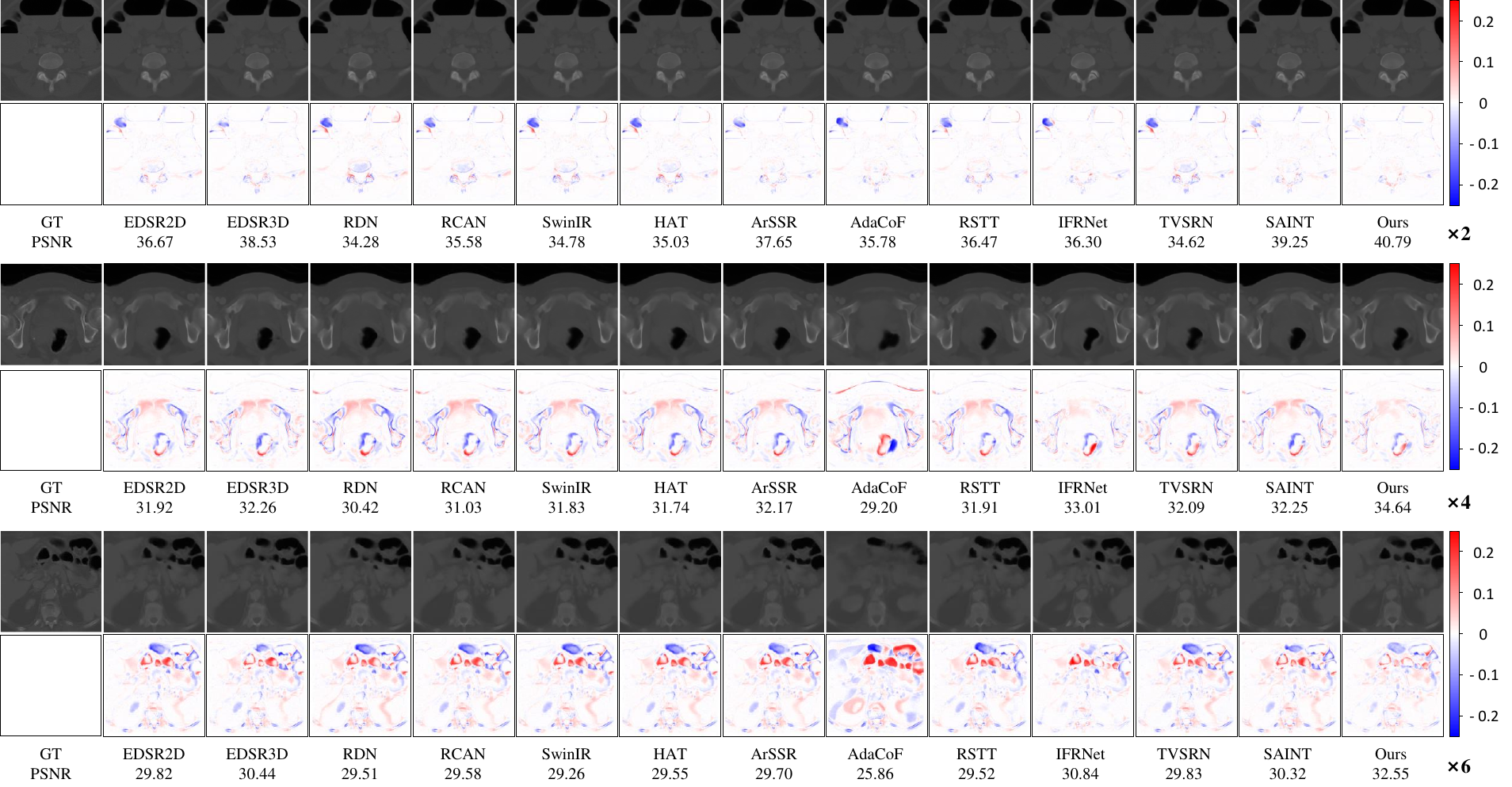}
	\caption{{Visual comparison against different methods on colon dataset. Every two rows, from top to bottom, represent the experimental results and {the error maps with the error bars} for $\times$2, $\times$4, and $\times$6 upscale factors, respectively. The lighter the color, the smaller the error.}}
    \label{fig:comparison_MSD}
\end{figure*}

As shown in Tables~\ref{tab:comparison_MSD}{-\ref{tab:comparison_IXI}}, we summarize the PSNR and SSIM from each view of {CT volume from MSD and MR volume from IXI} under $\times$2, $\times$4 and $\times$6 upscale factors, respectively. We compare our I$^3$Net with three types of methods: (1) super-resolution methods, including EDSR~\cite{lim2017enhanced}, RDN~\cite{zhang2018residual}, RCAN~\cite{zhang2018image}, SwinIR~\cite{liang2021swinir}{, HAT~\cite{chen2023activating}, ArSSR~\cite{wu2022arbitrary}} (similar to the pixel-wise interpolation), (2) video frame interpolation methods, including AdaCoF~\cite{lee2020adacof}, {RSTT~\cite{geng2022rstt}, }IFRNet~\cite{kong2022ifrnet} (similar to the slice-wise interpolation), and (3) typical slice interpolation methods TVSRN~\cite{yu2022rplhr}, SAINT~\cite{peng2020saint}. {The corresponding visualizations are shown in Fig.~\ref{fig:comparison_MSD} and Fig.~\ref{fig:comparison_IXI}.}

\begin{table}[!t]
\renewcommand\arraystretch{1.1}
\caption{Comparison with existing SR and interpolation methods under $\times 2$, $\times 4$ and $\times 6$ upscale factor, which is training on the colon, hepatic vessels and colon datasets, and testing on KiTS19.}
\label{tab:kids}
\centering
\footnotesize
\resizebox{1\linewidth}{!}{
    \begin{tabular}{|l|c|c| c|c |c|c|}
        \hline
        \multirow{2}{*}{Method} & \multicolumn{2}{c|}{$\times 2$} & \multicolumn{2}{c|}{$\times 4$} & \multicolumn{2}{c|}{$\times 6$} \\ \cline{2-7}
          & PSNR & SSIM$_a$  & PSNR & SSIM$_a$ & PSNR & SSIM$_a$ \\ \hline
        EDSR2D \cite{lim2017enhanced}   & 42.84 & 0.9769 & 35.62 & 0.9401 & 32.61 & 0.9173 \\
        EDSR3D \cite{lim2017enhanced}   & 43.16 & 0.9773 & 35.62 & 0.9397 & 32.47 & 0.9136 \\
        RDN~\cite{zhang2018residual}    & 42.05 & 0.9744 & 34.98 & 0.9370 & 30.97 & 0.9111 \\
        RCAN~\cite{zhang2018image}      & 42.04 & 0.9743 & 35.12 & 0.9369 & 31.81 & 0.9132 \\
        SwinIR~\cite{liang2021swinir}   & 42.41 & 0.9757 & 35.48 & 0.9394 & 32.15 & 0.9167 \\
        {HAT~\cite{chen2023activating}}   &{42.31}&{0.9758}&{35.53}&{0.9406}&{35.52}&{0.9178} \\
        {ArSSR~\cite{wu2022arbitrary}}    &{42.73}&{0.9767}&{35.58}&{0.9397}&{32.35}&{0.9143} \\
        AdaCoF~\cite{lee2020adacof}  & 42.28 & 0.9759 & - & - & - & - \\
        {RSTT~\cite{geng2022rstt}}     &{42.35}&{0.9749}&{35.30}&{0.9362}&{32.67}&{0.9137} \\
        IFRNet~\cite{kong2022ifrnet} & 42.36 & 0.9771 & 35.63 & 0.9440 & 32.74 & 0.9204 \\
        TVSRN~\cite{yu2022rplhr}     & 41.58 & 0.9702 & 35.16 & 0.9330 & 31.94 & 0.9065 \\
        SAINT~\cite{peng2020saint}   & 42.79 & 0.9766 & 35.60 & 0.9395 & 32.43 & 0.9117 \\ \hline
        \textbf{I$^{3}$Net (ours)}       & \textbf{43.54} & \textbf{0.9787} & \textbf{36.04} & \textbf{0.9444} & \textbf{32.95} & \textbf{0.9207} \\
        \hline
    \end{tabular} %
    }

\end{table}

\begin{table}[!t]
\renewcommand\arraystretch{1.1}
\centering
    \caption{Inference time (ms) comparisons.}

    \label{tab:abl_cost}
    \setlength{\tabcolsep}{4mm}{
    \begin{tabular}{|c|c|cc|}
        \hline
        Method & Inference Time  & PSNR  & SSIM$_\textrm{a}$  \\ \hline
        EDSR3D \cite{lim2017enhanced} & {122} & 32.71 & 0.9320 \\ 
        {HAT \cite{chen2023activating}} & {208} & {33.03} & {0.9341} \\
        {ArSSR \cite{wu2022arbitrary}} & {233} & {33.13} & {0.9355} \\
        {RSTT \cite{geng2022rstt}} & {275} & {32.94} & {0.9331} \\
        IFRNET \cite{kong2022ifrnet} & 168 & 33.88 & 0.9443 \\
        TVSRN \cite{yu2022rplhr} & 164 & 33.28 & 0.9355 \\
        SAINT \cite{peng2020saint}& 3,289& 32.76 & 0.9327 \\\hline
        \textbf{I$^3$Net (ours)} & \textbf{95}  & \textbf{34.42} & \textbf{0.9459} \\
        \hline
    \end{tabular}
    }
\end{table}

\begin{table}[t]
\renewcommand\arraystretch{1.1}
\centering
    \caption{Effectiveness of different modules in the whole network. All experiments are conducted using colon dataset with $\times$2 upscale factor.}
    \label{tab:abl_all}
    \setlength{\tabcolsep}{1mm}{
    
    \begin{tabular}{|ccc|cc|}
        \hline
        Intra-slice Branch & Inter-slice Branch & {Cross-view Block} & PSNR & SSIM$_\textrm{a}$ \\ \hline
          &  &  & 42.40 & 0.9841 \\ 
        \checkmark &  &  & 42.89 & 0.9851 \\ 
        \checkmark & \checkmark & & 43.72 & 0.9857 \\ 
        \checkmark & \checkmark & \checkmark & \textbf{43.90} & \textbf{0.9871} \\  
        \hline
    \end{tabular}
    }
\end{table}

\begin{table}[t]
\renewcommand\arraystretch{1.1}
\centering
    \caption{Performing slice synthesis from different views. We summarize the results of slice interpolation from coronal and sagittal views via super-resolution, results from the axial view via slice-wise interpolation, and results via fusing predictions from three views. All experiments are conducted using colon dataset with $\times$2 upscale factor. CVB stands for Cross-view Block.}
    \label{tab:diff_views}
    \setlength{\tabcolsep}{3mm}{
    \begin{tabular}{|c|c|cccc|}
        \hline
        \multicolumn{2}{|c|}{Method} & Coronal & Sagittal & Axial & Fuse  \\ \hline
        \multicolumn{2}{|c|}{RDN~\cite{zhang2018residual}}    & 40.83   & 40.80    & 41.21 & 41.50 \\ 
        \multicolumn{2}{|c|}{RCAN~\cite{zhang2018image}}   & 41.16   & 41.14    & 41.68 & 41.88 \\ 
        \multicolumn{2}{|c|}{EDSR~\cite{lim2017enhanced}}   & 41.52   & 41.57    & 42.40 & 42.73 \\ \hline
        \multirow{2}*{I$^{3}$Net}& w/o CVB & - &  -       & 43.72 & 43.69 \\ 
        ~                       & w CVB   & -  &  -       & \textbf{43.90} & -     \\
        \hline
    \end{tabular}
    }
\end{table}

\begin{figure}
    \centering
    \includegraphics[width=.5\textwidth]{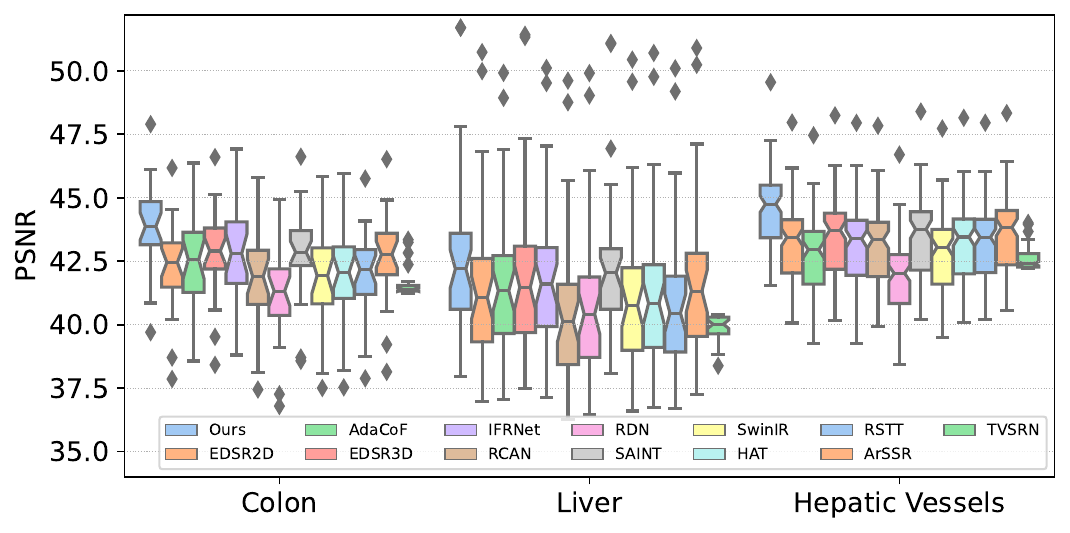}
    \caption{{Performance (PSNR) comparison in box plots of different methods using colon, liver and hepatic vessels datasets under $\times$2 upscale factor.}}
    \label{fig:boxplot_x2}
\end{figure}

\begin{figure*}[t]
	\centering
    \includegraphics[width=\linewidth]{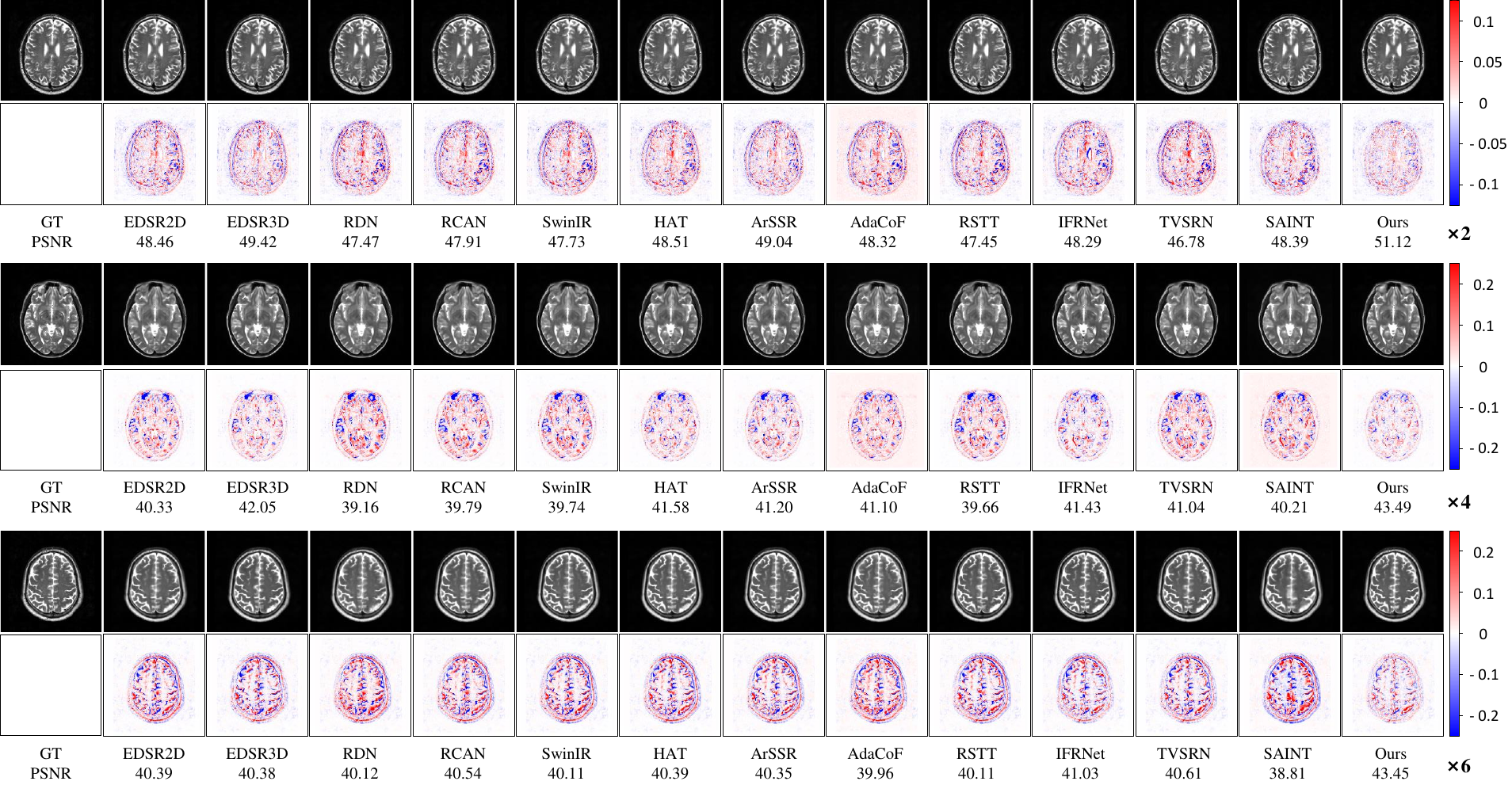}
	\caption{{Visual comparison against different methods on IXI dataset. Every two rows, from top to bottom, represent the experimental results and {the error maps with the error bars} for $\times$2, $\times$4, and $\times$6 upscale factors, respectively. The lighter the color, the smaller the error.}}
    \label{fig:comparison_IXI}
\end{figure*}

Tables~\ref{tab:comparison_MSD}{-\ref{tab:comparison_IXI}} show that our I$^{3}$Net is much better than those typical SR methods {based on convolution}{ (\emph{i.e.}, EDSR, RDN and RCAN) and some other SR methods based on transformer (\emph{i.e.}, SwinIR and HAT).} Compared with those typical SR methods, our I$^{3}$Net always performs best under different scale factors. Table~\ref{tab:diff_views} demonstrates that our I$^{3}$Net achieves higher PSNR of 2.51 dB, 2.04 dB, and 1.32 dB, respectively, compared to RDN, RCAN, and EDSR when performing slice interpolation from the axial view in the absence of cross-view blocks. Furthermore, these methods exhibit improved performance when fusing multiple views, whereas our I$^{3}$Net does not show a significant performance boost after the fusion strategy. This indicates that our I$^{3}$Net effectively extracts relevant information from the axial view, highlighting its efficacy.

As a 3D method, EDSR3D shows superior performance over some 2D methods under $\times$2 upscale factor. However, as the upscale factor increases, the performance of EDSR3D gradually declines compared to 2D methods. In comparison to the PSNR of EDSR3D, our I$^{3}$Net achieves an average improvement of 2.39\%, 4.39\% and 5.12\% {on MSD dataset} under the upscale factors of $\times$2, $\times$4, and $\times $6, respectively. The direct adaptation of the transformer-based SwinIR and {HAT} into slice interpolation does not have optimal performance.

{Another SR method, ArSSR, is based on INR. Indeed, ArSSR excels in performing arbitrary upsampling. However, as the upscale factor increases, the computation of ArSSR also escalates rapidly, while other regular SR methods and our I$^3$Net do not.}

Compared to methods for video frame interpolation, AdaCoF{, RSTT} and IFRNet have {mainly} achieved above-average performance results due to the similarity of the tasks{, especially when the through-slice resolution is low}. However, since AdaCoF was originally designed for $\times$2 upscaling, we do not include it in the comparison under the upscale factors of $\times$4 and $\times$6 in Tables~\ref{tab:comparison_MSD}{-\ref{tab:comparison_IXI}}. Nevertheless, its results are visualized in Fig.~\ref{fig:comparison_MSD}{and Fig.~\ref{fig:comparison_IXI}}, which shows that using AdaCoF directly under larger upscale factors makes it challenging to accurately predict the organ variation trends in the interpolated slices. {RSTT , without utilizing the optical flow, generally achieves lower performance compared to IFRNet.} In Fig. \ref{fig:comparison_MSD}{and Fig.~\ref{fig:comparison_IXI}}, the results of IFRNet show its capability to predict local information such as edges effectively. However, since IFRNet applies optical flow, it seems to overly rely on the content from adjacent slices, which results in better prediction of organ structures with minimal morphological changes between the preceding and succeeding slices. However, it struggles to accurately predict the scaling of organs.

\begin{figure}[t]
	\centering
    \includegraphics[width=\linewidth]{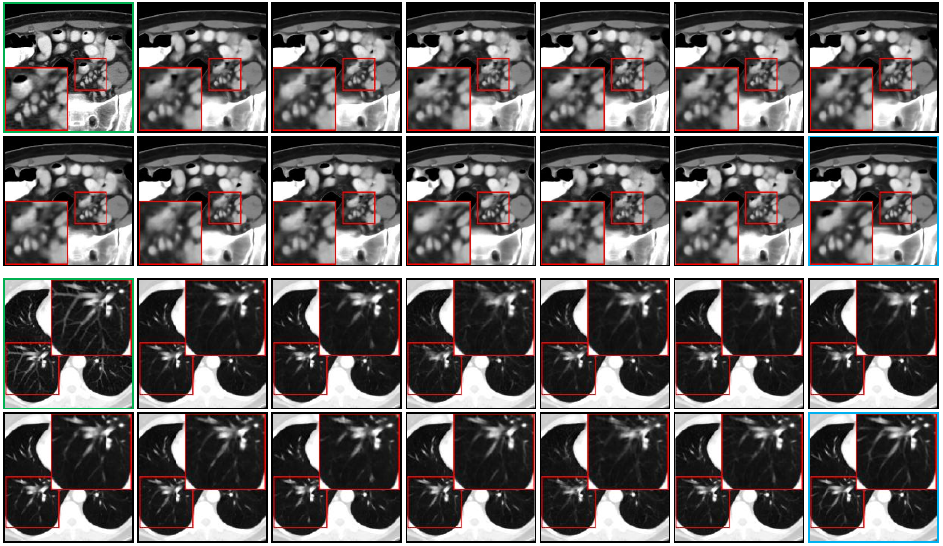}
	\caption{{The visualization results after applying HU windows. The window for the first two rows is set to [-125, +275], while the window for the last two rows is set to [-900, +100].} Green boxes and blue boxes are GT and our method, respectively. Other methods (\emph{i.e.}, EDSR2D, EDSR3D, RDN, RCAN, SwinIR, HAT, ArSSR, AdaCoF, RSTT, IFRNet, TVSRN and SAINT) are drawn in block boxes.}
    \label{fig:viual_set_window}
\end{figure}

Our I$^{3}$Net outperforms the pure transformer-based TVSRN, especially under lower upscale factors. In comparison to the PSNR, ours achieves an average improvement of 5.78\%, 3.54\% and 2.79\% {on MSD dataset} under the upscale factors of $\times$2, $\times$4, and $\times$6, respectively. SAINT only performs SR from coronal and sagittal views, without directly analyzing the axial view. As shown in Fig.~\ref{fig:comparison_MSD}{ and Fig.~\ref{fig:comparison_IXI}}, our I$^{3}$Net shows better performance in the prediction of organ boundaries. Fig.~\ref{fig:boxplot_x2} illustrates the superior performance of our I$^{3}$Net across all datasets, with substantial improvements. The p-values are lower than $1e^{-4}$, indicating significantly better performance compared to the second-best methods, IFRNet and SAINT.

{To enhance contrast and highlight differences, we apply HU windows to CT results shown in Fig.~\ref{fig:viual_set_window}. [-125,+275] is set for the abdominal region and [-900,+100] is set for the lung region. It can be observed that our method can reconstruct organs with greater clarity and accuracy, as well as capture complex variations of vessels.
}

To verify the robustness, we conduct the experiments on unseen KiTS19 data using the well-trained model on colon, hepatic vessels and liver. As indicated in Table~\ref{tab:kids}, {I$^3$Net still achieves the best results}.

\subsubsection{Inference Time Comparison}
We select {some} methods with the best overall performance and measure their speed when processing 4 input slices to synthesize 19 output slices. The inference time is computed on NVIDIA GeForce RTX 3090 GPUs. As shown in Table~\ref{tab:abl_cost}, I$^3$Net takes the lowest inference time while obtaining the best synthesis results.

\begin{figure}[t]
    \centering 
    \subfloat[\label{fig:high_freq_energy.sub.a}]{
            \includegraphics[height=.18\textwidth]{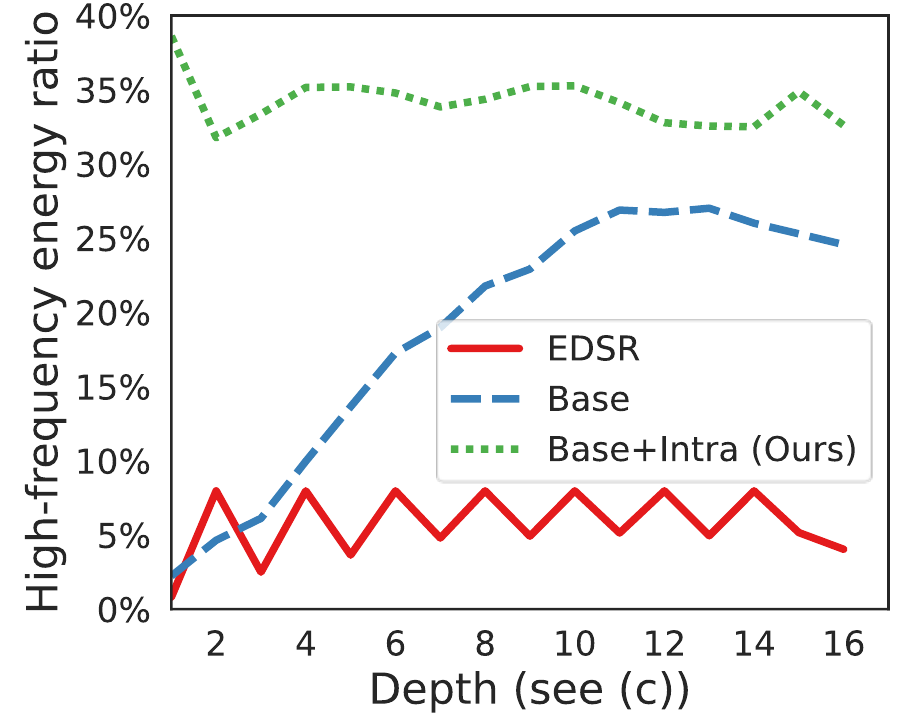}}
    \subfloat[\label{fig:high_freq_energy.sub.b}]{
            \includegraphics[height=.18\textwidth]{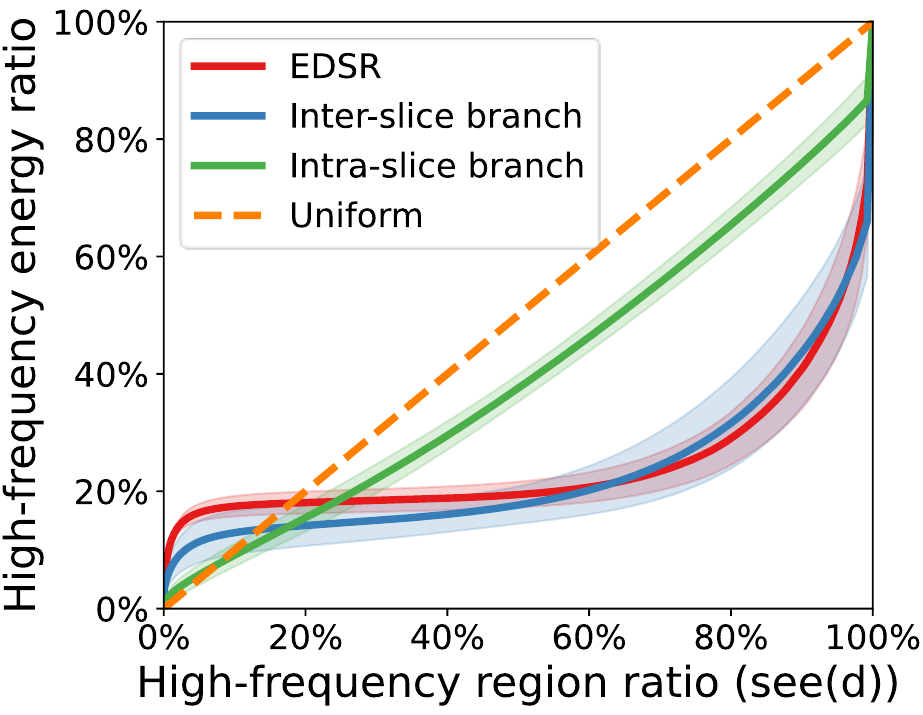}}
    \\
    \subfloat[\label{fig:high_freq_energy.sub.c}]{
            \includegraphics[width=.47\textwidth]{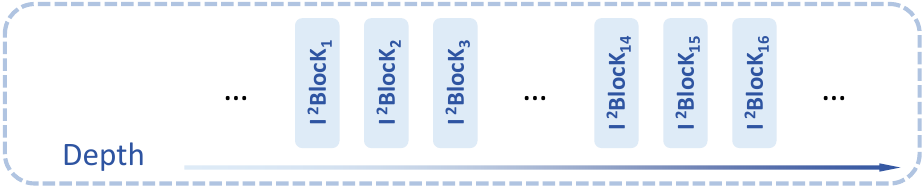}}
    \\
    \subfloat[\label{fig:high_freq_energy.sub.c}]{
            \includegraphics[width=.47\textwidth]{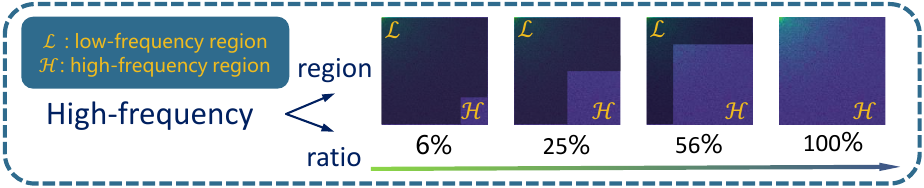}}
        
    \caption{{(a) shows the high-frequency energy across each stage of each method. ``Base'' only contains inter-slice branch in the I$^2$Block. ``+Intra'' means adding the intra-slice branch to I$^2$Block.} (b) {shows the proportion of the high-frequency energy ratio of each branch in the I$^2$Block and EDSR. ``Uniform'' has the equal intensity of every frequency band, showing {\textbf{IDEAL} equal learning ability} for high- and low-frequency bands (see Sec.~\ref{sec:intra}). The horizontal axis represents the range of the defined high-frequency region ratio.} (c) shows the horizontal axis of (a): increasing depth as the blocks (ResBlocks in EDSR and I$^2$Blocks in Base) are added. (d) shows the horizontal axis of (b): high-frequency region ratio. High-frequency energy ratio of (b) is computed as $\textstyle \sum_{(i,j)\in\mathcal{H}}I_{i,j}^{2} / \textstyle \sum_{(i,j)\in \mathcal{L} \bigcup \mathcal{H}}I_{i,j}^{2}$, where $I$ is the intensity of corresponding frequency.}
    \label{fig:high_freq_energy}
\end{figure}

\subsection{Ablation Study}

In this section, each component of our method is evaluated to demonstrate its effectiveness in improving the performance, which is shown in Table \ref{tab:abl_all}.
For different components, we analyze their respective effects in the following.

\subsubsection{Intra-slice Branch}
We conduct ablation experiments to analyze the influence of different components for our intra-slice branch. As shown in Table~\ref{tab:abl_intra_1}, without DCT operation, window partition + MLP results in 0.50 dB drop in PSNR. This is because only local information is learned. Moreover, Fig.~\ref{fig:high_freq_energy} (a) shows adding the intra-slice branch can result in increased focus on high-frequency contents. {Fig.~\ref{fig:high_freq_energy} (b) shows that when involving more frequency bands starting from the high frequency ($|\mathcal{H}|/|\mathcal{L}+\mathcal{H}|$ in Fig.~\ref{fig:high_freq_energy} (b)), the high-frequency energy ratio curve of the intra-slice branch is closer to that of the Uniform than the inter-slice branch and EDSR (a plain network). 
This shows our intra-slice branch ensures a more equal learning opportunity for different frequency bands.}
To further show our intra-slice branch \emph{i.e.}, learning high- and low-frequency independently, is a more effective method than other global context modeling methods, we replace the intra-slice branch with some typical non-local methods, including MAXIM~\cite{tu2022maxim}, CCNet~\cite{huang2019ccnet}, Twins~\cite{chu2021twins}, Swin~\cite{liu2021swin}. Table~\ref{tab:abl_intra_2} shows intra-slice branch has the best performance and achieves 0.44 dB higher than MAXIM, which is also based on MLP-Mixer.
Furthermore, an additional design is tested in Table~\ref{tab:abl_window}. We find increasing the size of the window to 16 has the best performance, yet excessively large window sizes result in a decline in performance.

\begin{table}[t]
\centering

\begin{minipage}{.23\textwidth}
    \renewcommand\arraystretch{1.1}
    \footnotesize
    \centering
    \setlength{\tabcolsep}{3pt}
        \caption{Effectiveness of each component in the intra-slice branch. All experiments are conducted using colon dataset with $\times$2 upscale factor.}
        \label{tab:abl_intra_1}
        \vspace{3mm}
		\begin{tabular}{|ccc|cc|}
                \hline
                DCT & Win & MLP& PSNR & SSIM$_\textrm{a}$ \\ \hline
                           &            &            & 42.36 & 0.9839 \\ 
                \checkmark &            & \checkmark & -     & -      \\ 
                           & \checkmark & \checkmark & 41.57 & 0.9817 \\ 
                \checkmark & \checkmark & \checkmark & \textbf{42.89} & \textbf{0.9851} \\
                \hline
		\end{tabular}
\end{minipage}\quad
\begin{minipage}{.23\textwidth}
    \renewcommand\arraystretch{1.1}    
    \footnotesize
    \centering
        \caption{Replacing intra-slice branch with typical non-local methods. All experiments are conducted using colon dataset with $\times$2 upscale factor.}
        \label{tab:abl_intra_2}

		\begin{tabular}{|c|cc|}
                \hline
                Method  & PSNR & SSIM$_\textrm{a}$ \\ \hline
                MAXIM   & 42.46 & 0.9841 \\ 
                CCNet   & 42.42 & 0.9842 \\ 
                Twins   & 41.95 & 0.9829 \\ 
                Swin    & 42.34 & 0.9837 \\
                \textbf{Ours}    & \textbf{42.89} & \textbf{0.9851} \\
                \hline
		\end{tabular}
\end{minipage}

\end{table}

\begin{table}[t]
\centering

\begin{minipage}{.23\textwidth}
    \renewcommand\arraystretch{1.1}
    \footnotesize
    \centering
        \caption{Different sizes of non-overlapping windows in the intra-slice branch. All experiments are conducted using colon dataset with $\times$2 upscale factor.}
        \label{tab:abl_window}

        \begin{tabular}{|c|cc|}
            \hline
            Win size & PSNR & SSIM$_\textrm{a}$ \\ \hline
            4           & 43.89 & \textbf{0.9871}  \\
            8           & 43.88 & \textbf{0.9871} \\
            \textbf{16} & \textbf{43.90} & \textbf{0.9871}  \\
            24          & 43.83 & 0.9870  \\
            32          & 43.66 & 0.9868  \\
            \hline
        \end{tabular}
\end{minipage}\quad
\begin{minipage}{.23\textwidth}
    \renewcommand\arraystretch{1.1}    
    \footnotesize
    \centering
        \caption{Different feature extraction strategies for the inter-slice branch. All experiments are conducted using colon dataset with $\times$2 upscale factor.}
        \label{tab:abl_inter}

        \begin{tabular}{|c|cc|}
                \hline
                Strategy  & PSNR & SSIM$_\textrm{a}$ \\ \hline
                Baseline   & 42.89 & 0.9839 \\ 
                Dilate       & 42.77 & 0.9848 \\ 
                Conv5         & 43.00 & 0.9853 \\ 
                S-T        & 43.66 & 0.9866 \\  
                \textbf{PC (Ours)}   & \textbf{43.72} & \textbf{0.9867} \\
                \hline
        \end{tabular} 
\end{minipage}

\end{table}

\begin{figure}[t]
    \centering
    \includegraphics[width=0.45\textwidth]{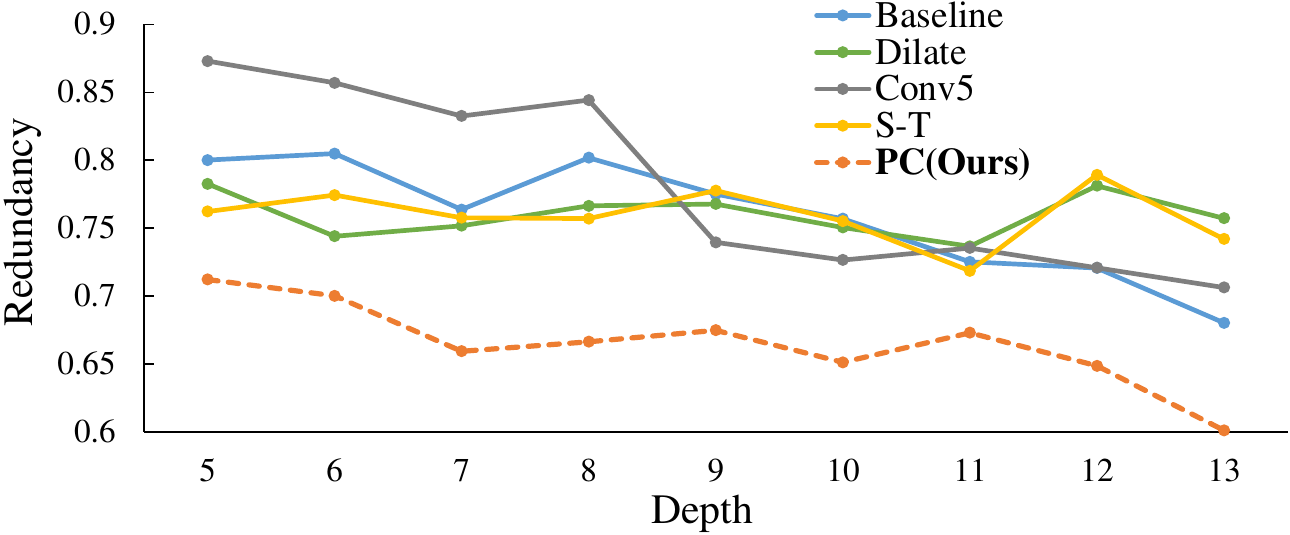}
    \caption{Feature redundancy vs. network depth on colon dataset using different strategies for the inter-slice branch.}
    \label{fig:redundancy}
\end{figure}

\subsubsection{Inter-slice Branch}
Instead of performing PixelUnShuffle, convolution, ReLU and PixelShuffle operations to extract enriched and diverse information between different slices, we also explore other feature extraction strategies. Given the feature map $\mathbf{Z} \in \mathbb{R}^{C \times H \times W}$, after applying PixelUnshuffle operation, we obtain the feature map $\mathbf{Z}^{pu} \in \mathbb{R}^{4C \times {H}/{2} \times {W}/{2}}$. Subsequently, convolution with a kernel size of 3 is performed. This process, convolution after PixelUnshuffle, is denoted by {PC}. The calculated pixels correspond to spatial locations in $\mathbf{Z}$ that are spaced 1 pixel apart, just like performing a convolution calculation in $\mathbf{Z}$ with a kernel size of 3 and a dilation of 1, which is denoted by {Dilate}. Due to the presence of dilation, adjacent points in $\mathbf{Z}$ will not be calculated together during {Dilate}, but they can be learned together across different channels during 
our {PC}. To ensure that adjacent pixels can be simultaneously activated while maintaining the same receptive field, we directly adopt a convolution with the kernel size of 5 to replace {Dilate}, representing this process as {Conv5}. To demonstrate the superior ability to preserve original information by PixelUnshuffle and PixelShuffle, we use convolution with stride for downsampling and transpose convolution for upsampling, which process is represented as {S-T}. It can be observed from Table~\ref{tab:abl_inter} that our {PC} achieves the best performance, surpassing the performance of the parameterized upsampling and downsampling method {S-T} by 0.06 dB, and it outperforms other methods by at least 0.7 dB. Furthermore, according to the redundancy reduction principle, high feature redundancy limits the generalization of neural networks \cite{zbontar2021barlow}. We show our {PC} leads to the lowest feature redundancy, compared with other strategies in Fig.~\ref{fig:redundancy}. Besides, with the increasing of the network depth, our {PC} exhibits a clear decreasing trend.

\begin{figure}[t]
    \centering
    \includegraphics[width=0.45\textwidth]{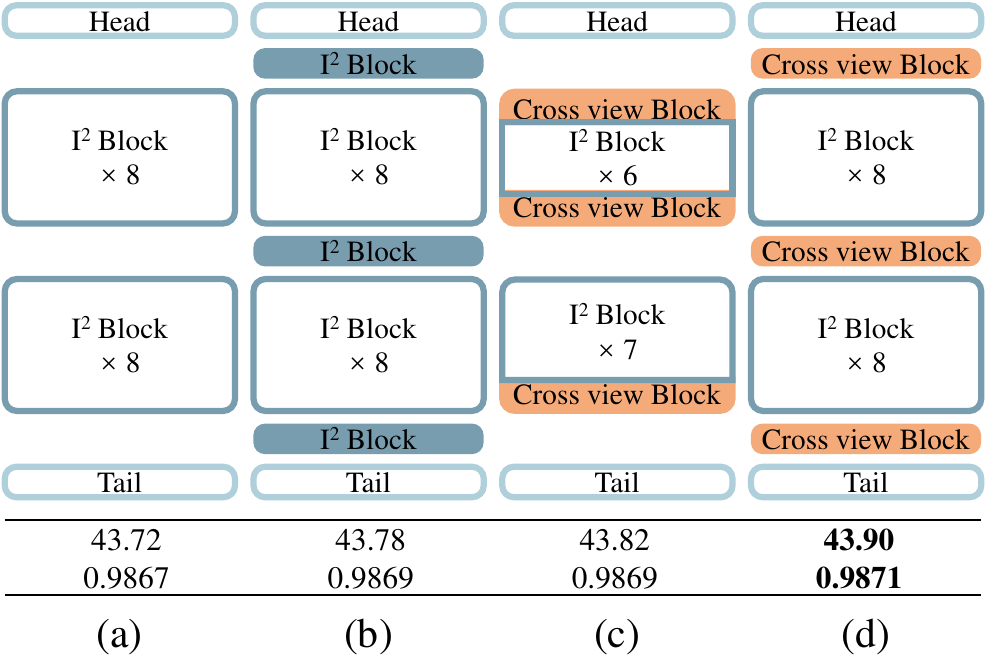}
    \caption{Effectiveness of Cross-view Block at different positions in the network. The last two rows of metrics represent the PSNR and SSIM$_\textrm{a}$ for different strategies, respectively.}
    \label{fig:abl_cross}
\end{figure}

\subsubsection{Cross-view Block}
To verify the effectiveness of cross-view block, we conduct experiments by placing cross-view blocks at different positions in the network. The specific strategy for the experiments is illustrated in Fig.~\ref{fig:abl_cross}, with the corresponding results presented below. In Fig.~\ref{fig:abl_cross}, strategy (a) shows the backbone of the network consists of 16 I$^2$ blocks, serving as the baseline. Strategy (d), our method, inserts three cross-view blocks into the network. To show that the performance enhancement is not due to an increase in the parameter numbers, we implement strategy (b), which adds three extra I$^2$ blocks. Additionally, strategy (c) replaces three I$^2$ blocks with three cross-view blocks, which preserves the overall number of blocks compared with (a). We observe performance gains between (a) and (c), (b) and (d), showing the effectiveness of our cross-view block. With fewer parameters, strategy (c) even outperforms strategy (b) in terms of PSNR by 0.04 dB, which indicates that the cross-view block effectively integrates the features from sagittal and coronal views. Our cross-view block finally leads to an increase of 0.18 dB in PSNR.

\begin{figure}[t]
    \centering
    \includegraphics[width=\linewidth]{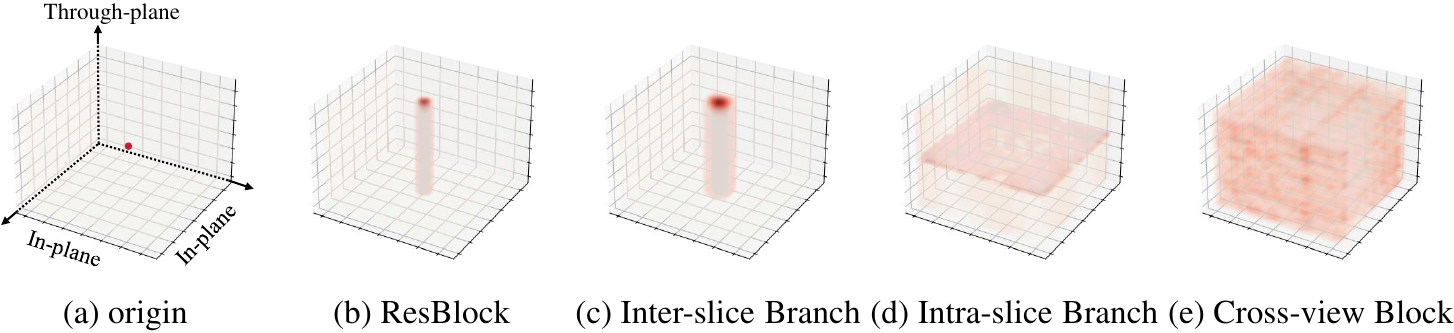}
    \caption{{Reception field of each module with the origin as the input data. ResBlock is the baseline for comparison.}}
    \label{fig:reception_field}
\end{figure}

\begin{figure}[t]
    \centering
    \includegraphics[width=\linewidth]{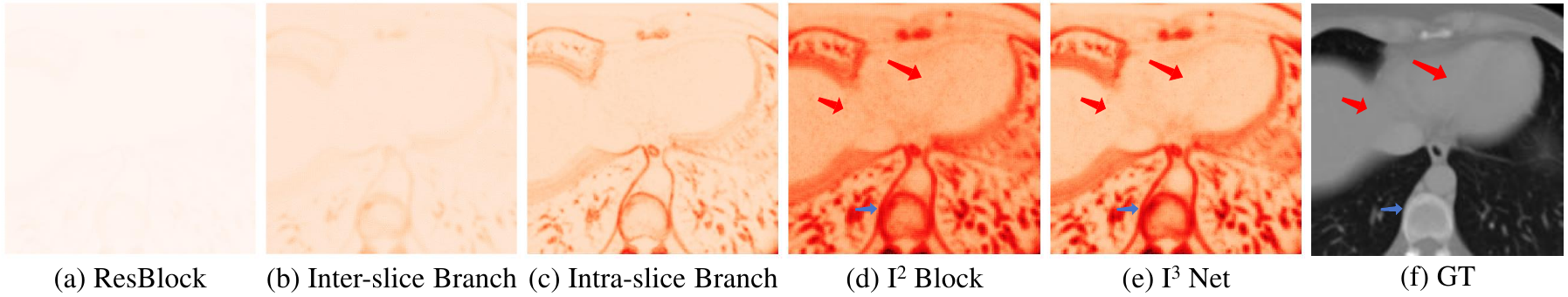}
    \caption{{Feature variations after passing through the module. ResBlock is the baseline for comparison.}}
    \label{fig:feature_difference}
\end{figure}

\subsubsection{Number of Each Module}
{To explore the impact of the I$^2$Block and Cross-view Block on performance enhancement, we conduct experiments on different numbers of these two modules. Table~\ref{tab:abl_num_i2block} illustrates digging deeply into axial-view (increasing the number of I$^2$Blocks) helps improve the performance. Fusing through-plane information can assist in enhancing the performance. However, simply increasing the number of cross-view blocks does not guarantee an effective improvement in performance as shown in Table~\ref{tab:abl_num_cvb}. The appropriate number of Cross-view Blocks can provide valuable through-plane information. Due to the typically lower through-plane resolution compared to the in-plane resolution, excessive integration of through-plane information will cause interference.}

\begin{table}[b]
\renewcommand\arraystretch{1.1}
\centering
\caption{{Different number of I$^2$Blocks. All experiments are conducted using colon dataset with $\times$2 upscale factor.}}
        \centering
        \begin{tabular}{|p{20mm}|ccccc|}
                \hline
                \centering I$^2$Block & 8 & 12& 16 & 20 & 24 \\ \hline
                \centering w CVB   &  43.52 & 43.76 & 43.90 & 43.95 & 44.02 \\ 
                \centering w/o CVB &  43.38 & 43.62 & 43.72 & 43.72 & 43.84 \\  \hline
                \centering $\Delta$ & +0.14 & +0.14 & +0.18 & +0.23 & +0.18 \\
                \hline
		\end{tabular}
\label{tab:abl_num_i2block}
\end{table}

\begin{table}[b]
\renewcommand\arraystretch{1.1}
\centering
\caption{{Different number of Cross-view Blocks. All experiments are conducted using colon dataset with $\times$2 upscale factor.}}

\subfloat{
		\begin{tabular}{|p{20mm}|ccccc|}
                \hline
                \centering CVB  & 0 & 1 & 3 & 5 & 9 \\ \hline
                \centering PSNR & 43.72 & 43.85 & 43.90 & 43.86 & 43.83 \\ 
                \hline
		\end{tabular}
        }
\label{tab:abl_num_cvb}
\end{table}

\subsubsection{Complementary to Each Module}
{In the previous subsections, we have discussed the individual characteristics of each module. We then demonstrate the complementarity among these modules to enhance the overall network performance. As shown in Fig.~\ref{fig:reception_field}, each module focuses on different aspects. Compared to the baseline ResBlock, inter-slice branch has a larger local reception field through all channels. However, it still fails to capture global information within the plane. intra-slice branch, due to its property of equal learning for different frequency bands, is capable of extracting global information effectively, while it primarily focuses on in-plane information from neighbouring channels. Thus, these two branches complement each other by compensating for the in-plane and through-plane information, respectively. The reception field of the Cross-view Block indicates its capability to acquire anatomical information from the coronal and sagittal views, providing different perspectives to assist the inter- and intra-slice branches.}

{To provide an intuitive understanding of the impact of each module, we show the feature variation after passing through each module, as shown in Fig.~\ref{fig:feature_difference}. Obviously, inter-slice branch appears to perceive more effective information than ResBlock, but it lacks some details. On the other hand, intra-slice branch can pay more attention to high-frequency details. The integration of these two branches further ensures the overall feature extraction. The inclusion of Cross-view Block assists in feature adjustment, alleviating unnecessary variation in smooth regions (\emph{e.g.}, red arrows shown in Fig.~\ref{fig:feature_difference}) and highlighting regions with rapid and high-frequency variation (\emph{e.g.}, the blue arrow shown in Fig.~\ref{fig:feature_difference}), which leads to an improved contrast in feature variation.}

\begin{table}[t]
\renewcommand\arraystretch{1}
\centering
    \caption{{PSNR results of two contrastive learning stratgies with different wight $w$.}}
    \label{tab:contrastive_learning}
    \setlength{\tabcolsep}{3mm}{
    \begin{tabular}{|p{5mm}|ccccc|}
        \hline
        \centering $w$ & 0     & 0.05   & 0.10   & 0.20 & 0.50 \\ \hline
        \centering $\mathcal{L}_{c,1}$   & 43.90 & 43.88 & 43.92 & 43.88 & 43.75 \\ 
        \centering $\mathcal{L}_{c,2}$   & 43.90 & 41.38 & 41.04 & 40.73 & 40.18 \\ 
        \hline
    \end{tabular}
    }
\end{table}

\subsubsection{Optimization}
{Intuitively, neighboring patches often share similar semantic information~\cite{yun2022patch}. To fully exploit the prior of similarity among adjacent slices, we investigate the impact of contrastive learning on medical slice synthesis. Following~\cite{yi2021contrastive}, we design a contrastive loss $\mathcal{L}_{c,1}$ by treating neighboring slices of predicted results as positive samples and distant slices as negative samples, then pull positive samples closer and push negative samples farther away. The results of applying $\mathcal{L}_{c,1}$ with different weights are presented in Table~\ref{tab:contrastive_learning}, illustrating that it has almost no effect, even leads to a decline in performance. We also design a dense contrastive loss $\mathcal{L}_{c,2}$ in the same way, by selecting neighboring channels in the feature volume from I$^2$Block as positive samples and distant channels as negative samples. However, it results in a decline in performance as shown in Table~\ref{tab:contrastive_learning}. Due to significant variations among adjacent slices and potential similarities among distant slices in some cases, applying contrastive loss may provide erroneous guidance. Thus, we only adopt $L_1$ Loss following prior works~\cite{peng2020saint,yu2022rplhr,lim2017enhanced}.}

\section{Conclusion}
When synthesizing thicker slices of the anisotropic CT volumes, we observe that directly adopting slice-wise interpolation from the axial view outperforms super-resolution from other views, and it is a more concise way compared to fusing from multiple views. In this paper, we propose an I${^3}$Net to explore the {in-plane} information from the axial view with high resolution and get compensation from other views for medical slice synthesis. {Inter-slice branch takes high {in-plane} resolution information to enrich {through-plane} features. Intra-slice branch ensures long-range dependencies equal learning of every frequency bands, while preventing spatial discretization. To fully explore CT volumes, the cross-view block is proposed to compensate for information from omni directions}. Experiments on within-datasets and cross-datasets verify the effectiveness of our method.





\bibliographystyle{IEEEtran}
\bibliography{reference.bib}

\end{document}